\documentclass[letterpaper, 10pt, conference]{ieeeconf}
 \IEEEoverridecommandlockouts
 \usepackage{tikz}
 \usepackage{siunitx}
\usepackage{float}
\usepackage{pbox}
\usepackage{epstopdf}
\usepackage{setspace}
\usepackage{adjustbox}
\usepackage[colorlinks=true,linkcolor=blue]{hyperref}
\usepackage{hhline}
\usepackage{url}
\usepackage{amsmath,amsfonts,amssymb}
\usepackage{mathtools,amssymb}
\usepackage{graphicx}
\DeclareGraphicsExtensions{.pdf,.png,.jpg}
\usepackage{microtype}
\usepackage{algorithm}
\usepackage{algorithmic}
\usepackage{adjustbox}
\usepackage{color}
\usepackage{xcolor}
\usepackage{multirow}
\usepackage{graphicx}
\usepackage{flushend}
\usepackage{subfigure}
\usepackage[font=small,skip=0pt]{caption}

\usepackage{amssymb,mathtools}
\usepackage{xfrac}
\usepackage{enumerate}
\usepackage{hyperref}
\usepackage{cite}

\newtheorem{remark}{Remark}

\title{\LARGE \bf
Extending life of Lithium-ion \textcolor{black}{battery systems by \textcolor{black}{embracing}} heterogeneities via an optimal control-based active balancing strategy}

\author{
Vahid Azimi, Anirudh Allam,  Simona Onori, \textit{Senior Member, IEEE}
\thanks{Vahid Azimi, Anirudh Allam, Simona Onori are with the Energy Resources Engineering Department, Stanford University, Stanford, CA 94305 USA, \mbox{\{vazimi,\ aallam,\ sonori\}@stanford.edu}. (\textit{Corresponding author: Simona Onori.})}
}
\hypersetup{draft}
\begin{document}

\maketitle
\thispagestyle{empty}
\pagestyle{empty}

	\begin{abstract}

\textcolor{black}{\textcolor{black}{This paper formulates and solves a multi-objective fast charging-minimum degradation optimal control problem (OCP) for a lithium-ion battery module made of series-connected cells equipped with an active balancing circuitry. The cells in the module are subject to heterogeneity induced by  manufacturing defects and non-uniform operating conditions.} Each cell is expressed via a coupled nonlinear electrochemical, thermal, and aging model and the direct collocation approach is employed to transcribe the OCP into a nonlinear programming problem (NLP).
	The proposed OCP is formulated under two different schemes of charging operation: (i) same-charging-time (OCP-SCT) and (ii) different-charging-time (OCP-DCT).  The former assumes simultaneous charging of all cells irrespective of their initial conditions, whereas the latter allows for different charging times of the cells to account for heterogeneous initial conditions.
	%
	\textcolor{black}{The problem is solved for a module with two series-connected cells with intrinsic heterogeneity among them in terms of state of charge and state of health.} Results show that the OCP-DCT scheme provides more flexibility to deal with heterogeneity, boasting of lower temperature increase, charging current amplitudes, and degradation. \textcolor{black}{Finally, comparison with the common practice of constant current (CC) charging over a long-term cycling operation shows that promising savings, in terms of retained capacity, are attainable under the both control (OCP-SCT  and OCP-DCT) schemes.}}
\end{abstract}


\section{Introduction} \label{section.introduction}
\textcolor{black}{Lithium-ion batteries (LIBs) are the enabling technology to ensure a sustainable future thanks to their
	high cell voltage, high energy and power density, low memory effect,
	long life, and increasingly reduced  cost ~\cite{Li1}}. 
They have been  extensively utilized in a wide range of applications including  microgrids, consumer electronics, and  Electric Vehicles (EVs)~\cite{App3,App4,App5}. 
\textcolor{black}{Consumer acceptance of battery-powered devices is highly dependent on their fast-charging ability while maintaining a  safe and long-running operation. In EVs today, constant-current (CC) charging is used,  where the charger supplies a relatively uniform current, regardless of the battery \textcolor{black}{State of Charge (SOC)} or temperature \cite{CCCV1,CCCV2}.
	Batteries used in EVs consist of a large number of cells connected both in series and parallel. Variations in the parameters of individual battery cells, such as capacity mismatch, impedance, and operating temperature, are deemed to expand throughout the life of the device. One of the tasks of the  Battery Management System (BMS) is to provide cell balancing functionality by measuring and comparing the states of all cells after each charging cycle.
	Recent advances in battery life management have come from advanced BMS strategies that rely on battery models around which estimation and optimization strategies are designed. 
	Most of the recent battery control/optimization literature, though,  has focused on single cell operation under fast charging. Methods to optimizing longevity under fast charging operation for single cells have been proposed based on  model predictive control (MPC)~\cite{Zou2018, OptCellMPC3},  nonlinear programming (NLP)~\cite{OptCellGP1, Lam2021} and  Control Vector Parameterization (CVP) \cite{Pathak_2017}  either using equivalent circuit models or electrochemical models.
	However, the problem of battery fast charging while preserving its health is a pack-level challenge that needs to be tackled as such. }

\subsection{Motivation and Related Literature}
\textcolor{black}{A battery pack consists of individual cells, which are organized into modules made of cells connected in series/parallel.
Results obtained for single cells cannot be extrapolated or generalized to the module/pack level due to the loss of modularity in the system~\cite{allam2015battery}. 
A fundamental characteristic of an interconnected battery system (module and pack) is that heterogeneity in the parameters \textcolor{black}{within the series/parallel cells} is inevitable due to manufacturing and operating conditions which, \textcolor{black}{if not monitored or corrected on time, could hinder the performance and longevity of the battery system during operation \cite{Beck2021}}. 
\textcolor{black}{Manufacturing-induced heterogeneities, such as capacity and impedance of single cells, are deemed to be exacerbated over time, and at the same time, end up being the cause of differences in temperature, SOC, depth of discharge (DOD) and charging rate ~\cite{Lit1,Lit2,Todeschini2012}.}}
%
\textcolor{black}{For example, voltage and charge imbalances limit the charge/discharge capabilities of the pack, posing limitations on pack-level performance and causing temperature imbalance, which is known to accelerate battery pack aging \cite{Allam2019}}.

\textcolor{black}{Battery equalization methods are employed to bring the cells in a pack to the same voltage/SOC. \cite{GALLARDOLOZANO2014}.  These  methods fall into two main categories: passive and active balancing. In passive methods - for example, in the form of a fixed shunting resistor - no active control is used to balance the cells and the excess energy from the high SOC cells is dissipated until the charge matches the lower SOC cells in the pack.  
	Active balancing methods, on the other hand, offer more flexibility in equalizing the energy of each cell in the pack \cite{GALLARDOLOZANO2014} and rely on active control strategies. 
	\textcolor{black}{It is worth mentioning that in the literature, there is a lack of consensus as to what is interpreted as an active or passive balancing framework. In some cases, energy storing and redistributing components such as DC/DC converters are considered to constitute an active balancing circuit, and in other cases, the presence of a control strategy to balance the cells (either through switching shunt resistors, transistors, or DC/DC converters) is considered to constitute an active balancing circuit. }	
		 For example, in \textit{cell-bypass} active methods, implemented either via shunt resistor or shunt transistor method \cite{GALLARDOLOZANO2014}, the current of each cell is bypassed whenever the cell voltage reaches the admissible upper limit by means of a switch in series with a resistor or a transistor, respectively.  In the \textit{cell-to-cell} methods, in the form of, for example, bypass DC/DC converters \cite{Zane2015}, the extra energy stored in the most charged cells is transferred to the least charged cells. Alternatively, balancing and complete cell-bypassing can be achieved by a module-integrated distributed battery system architecture \cite{li2016module}, wherein each cell in the module is individually managed by the modular converter without the need for equalization circuits.   
		 \textcolor{black}{The proposed work falls in the category of active balancing, in accordance with \cite{GALLARDOLOZANO2014}, since an optimal controller is proposed to actively switch the shunt resistors in cell-bypass balancing methods or switch DC/DC converters in cell-to-cell balancing methods.}
}

\textcolor{black}{While hardware strategies to enable active balancing are in place, scant attention has been paid to synthesizing optimization-based control strategies for battery pack/module. The impact of different balancing strategies on cell-to-cell variations, in terms of SOC, maximum capacity, and resistance, is addressed in \cite{Docimo2017}, where a formal framework based on linearized electrochemical dynamics and multivariable control theory is used to 1) show that voltage balancing fails to eliminate capacity and resistance imbalance between cells, and 2) design a strategy that is able to eliminate charge, capacity and resistance imbalance within the lifespan of the pack.  In \cite{Altaf2017}, an electrothermal control scheme is  devised for load management of a battery module for on-board vehicle operation to tackle charge and temperature imbalances by leveraging constrained linear quadratic model predictive control. In \cite{Docimo2019} charge imbalance and temperature imbalance are also tackled  by using a formal framework based on MPC to obtain insights on how temperature imbalance can be controlled through an average current. A simplified linear parameter varying model is developed to represent charge and temperature imbalance. 
In \cite{Pozzi2020}, SOC imbalance in series-connected cells is controlled via a nonlinear model predictive control scheme upon proper simplifications of the electrochemical battery dynamics and insights on  an easily implementable power supply scheme are provided.}

\subsection{Main Contributions}
\begin{figure*}\centering
	\includegraphics[scale=0.6]{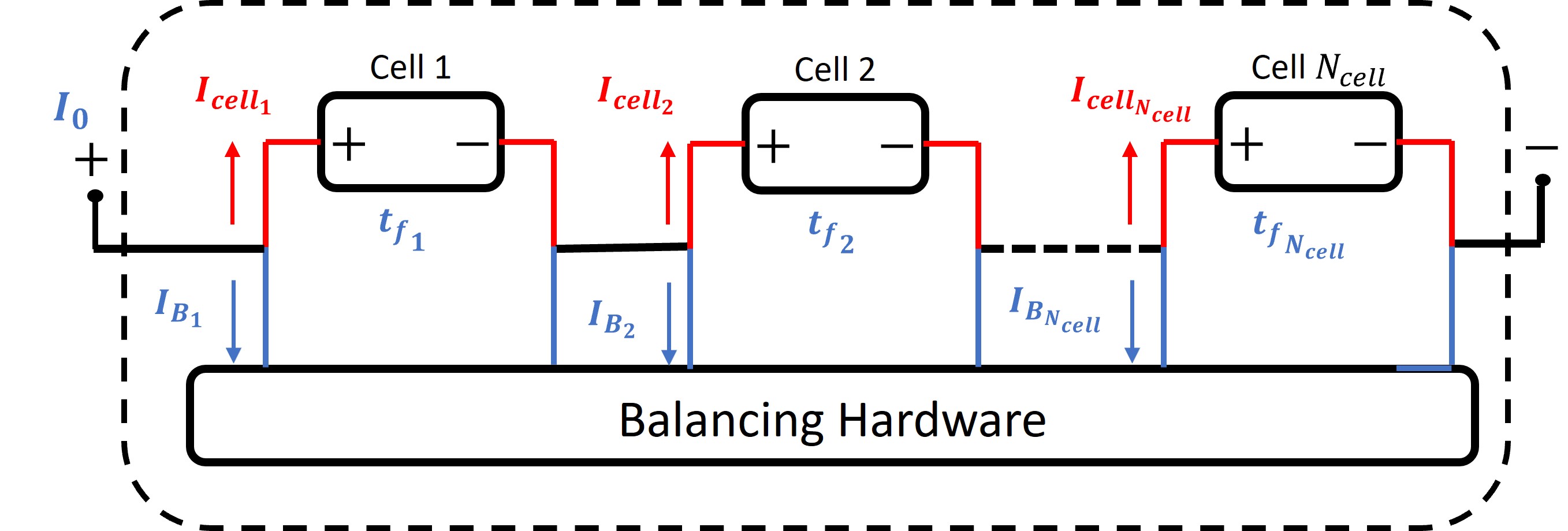}
		\vspace{0.5em}
	\caption{\textcolor{black}{Battery module with $N_{cell}$ series-connected cells, where each cell is connected to a balancing hardware. The current variable $I_0$ refers to the module current, variables $I_{cell_1}, \hdots I_{cell_{N_{cell}}}$ are the current magnitude through the cells, variables $I_{B_1}, \hdots I_{B_{N_{cell}}}$ are the balancing currents, and time variables $t_{f_1}, \hdots t_{f_{N_{cell}}}$ are the charging time associated with each cell.}}
	\label{fig.Pack4}
\end{figure*}

\textcolor{black}{In this paper, the system under investigation is a \textcolor{black}{LIB} module of $N_{cell}$ series-connected cells (see Fig.~\ref{fig.Pack4}), where each cell is connected to an active balancing hardware, which could be either as simple as an active shunt resistor or shunt transistor method or a more sophisticated hardware \textcolor{black}{such as} bypass DC/DC converters.\footnote{The specific hardware design is outside the scope of this paper. The reader can refer to \cite{GALLARDOLOZANO2014} for different active hardware balancing solutions.}
	For the given system, we address the problem of designing an optimization-based control strategy that controls individual cells to achieve fast charging while guaranteeing minimum degradation of the pack to be implemented in an active balancing hardware. In Fig.~\ref{fig.Pack4}, the current of the $k^{th}$ cell is given by $I_{cell_k}=I_0-I_{B_k}$, where $I_0$ is the module current  and $I_{B_k}$ is the current absorbed by the balancing hardware associated with $k^{th}$ cell. 
	Battery pack life optimization is achieved by controlling each individual cell \textcolor{black}{while embracing heterogeneities} in terms of state and parameters - due to either/both manufacturing defects or/and non uniform operating conditions. The formulated OCP will ultimately implement SOC balancing along with State of Health
	(SOH)-aware balancing by tackling the cell-to-cell heterogeneity. \textcolor{black}{The optimal control is multi-objective in nature to face the conflicting objectives of minimum time of charge ($t_{f_k}$) under minimum degradation by optimizing  the current profiles.}
	~\footnote{High C-rate currents would charge the battery faster at the expense of faster growth of Solid Electrolyte Interphase layer (SEI), causing capacity and power fade.}}

\textcolor{black}{Cells in the module are modeled via coupled nonlinear Partial Differential Equations (PDEs), Ordinary Differential Equations (ODEs), and Differential Algebraic Equations (DAEs)} describing the electrochemical, thermal, and aging dynamics~\cite{modelT}. \textcolor{black}{The Single Particle Model (SPM) is employed to model the electrochemical dynamics, a lumped two-state thermal model with cell-to-cell heat transfer terms is used to derive the core-cell temperature from ambient temperature, and finally aging is modeled through the growth of SEI layer on the negative electrode.}

Within the framework adopted, the cell is a multi-time scale system in which thermal dynamics acts as a fast dynamics, the electrochemical dynamics \textcolor{black}{as the semi-slow dynamics, and aging dynamics as the slow dynamics~\cite{Allam2019}}.
\textcolor{black}{The nonlinear and multi-time scale nature of the cell dynamics are retained in the formulation and solution of the multi-objective optimization problem addressed in this work
	\textit{and to the best of our knowledge, to date, no study has addressed such a problem using the high fidelity multi-time scale battery model.}} 
Noting the fact that aging dynamics \textcolor{black}{includes} SEI layer growth and solvent concentration, \textcolor{black}{where the former is a low-dimensional slow variable and the latter is a high-dimensional one}. 
\textcolor{black}{The computational burden imposed by 
the high fidelity dynamics at different time scales has led to the design of a surrogate model} to \textcolor{black}{approximate the high-dimensional slow dynamics (solvent concentrations) as a function of cell current and ambient temperature}. 

To solve the optimization problem, the direct collocation approach~\cite{DC1} is utilized to transcribe the OCP to a NLP~\cite{NLP} by parameterization of the system states and inputs, and charging times. The interior point solver IPOPT~\cite{IPOPT} is then used to solve the NLP problem \textcolor{black}{while the optimality of the solution is discussed using} the the Karush-Kuhn-Tucker (KKT) conditions (first-order necessary conditions). 
%
%
The OCP is formulated under two different schemes: (a) \textit{same-charging-time} (OCP-SCT) and (b) \textit{different-charging-time} (OCP-DCT).
To confirm the soundness of the proposed OCP-SCT and OCP-DCT schemes, simulation studies are carried out on an illustrative example \textcolor{black}{of} a battery module with two series-connected cells, each equipped with a active balancing hardware. 
The performance and robustness of the proposed schemes is shown under perturbation of parameters in terms  initial SOC and
initial  SOH imbalances (through variation in the initial SEI layer growth state).

\textcolor{black}{This paper extends on the initial investigation proposed in \cite{Azimi21} in that  1) it contains the description of the surrogate model used to capture the solvent diffusion dynamics of the SEI layer growth model, 2) characterizes the time scale difference of the LIB dynamics, 3) provides ample simulation scenarios of the two
optimization schemes for an effective and exhaustive comparison of the two and 4) provides a comparative study of the two charging scenarios with the traditionally used CC charging protocol.
\textcolor{black}{The main takeaways and recommendations from the proposed study are provided in the pursuit of a novel life-extension optimization charging strategy that embraces cell-to-cell heterogeneities by combining advanced optimization algorithms over multi-scale high fidelity models using active balancing hardware setup.}}


\subsection{Outline}
The organization of the paper is as follows.
\hyperref[section.mo] {Section~\ref*{section.notations}} lists the notations used in the paper. \textcolor{black}{{Section~\ref*{section.mo}} presents the mathematical model for cells and battery module. }
\hyperref[section.ps]{Section~\ref*{section.ps}} describes the problem statement. \hyperref[section.cont]{Section~\ref*{section.cont}} formulates the proposed optimal control methodology.
\hyperref[section.Sim]{Section~\ref*{section.Sim}} presents the simulation results.
\hyperref[section.Conc]{Section~\ref*{section.Conc}} presents the discussion and conclusion.


\section{Notations}\label{section.notations}
The following notations are used in the paper. 
\begin{itemize}
	\item Given a real $n$-dimensional vector $x$ with initial and final values $x(t_0)$ and $x(t_f)$ ($t_0$ and $t_f$ are the initial and final times), $\Delta x=\frac{|x(t_f)-x(t_0)|}{x(t_0)}\times 100\%$ is the percentage deviation of $x$ with respect to its initial value.
	\item Given the continuously differentiable function $f(x)$, $\nabla f(x)$ is the gradient of $f(x)$ with respect to $x$.
	\item The subscript $j\in[n; p]$  stands for the
	cell domain (e.g. $n$ = anode and $p$ = cathode)
	\item The subscript $i$ refers to
	the discretization grid position when converting PDEs to
	ODEs via Finite Difference Method (FDM) in solid electrodes and SEI layer spatial dimensions;
	\item The superscript $k$ represents the cell position within the series-connected module. 
\end{itemize}


\section{Battery Module Model} \label{section.mo}

This section presents the model \textcolor{black}{of} the LIB module with $N_{cell}$ series-connected cells.
Each cell is equipped with an active balancing circuitry that provides a practical way to reroute the current flowing in each cell, and that is used as an  extra degree of freedom to the optimal controller. 
\textcolor{black}{The nomenclature used in this section is listed at the end of the paper in Table \ref{table.ESPM_nomenclature}}.

\subsection{Cell electrochemical model}

\textcolor{black}{The SPM used to model the cell electrochemical dynamics assumes that each electrode is a single spherical particle and that the concentration gradient in the electrolyte}
phase is uniform, hence   the diffusion electrolyte dynamics can be neglected. SPM is described by two governing PDEs - one for each electrode - representing the mass conservation in the solid phase \textcolor{black}{through} 
Fick’s law
\begin{align}\label{eq:PDE_solid}
\frac {\partial c_{s,j}} {\partial t}= \dfrac{D_{s,j}(T_c)}{r^{2}} \dfrac{\partial}{\partial r} \left [r^{2} \dfrac{\partial c_{s,j}} {\partial r} \right] \qquad j\in[n,p]
\end{align}
associated with the \textit{Neumann boundary conditions} at the center and surface of the spherical particle given by
\begin{align}\label{eq:PDE_solidBound} \nonumber
\frac {\partial c_{s,j}} {\partial r}\Bigr|_{\substack{r=0}} &= 0 \\  
\frac {\partial c_{s,j}} {\partial r}\Bigr|_{\substack{r=R_{s,j}}}&=  \frac{\pm I_{cell} }{D_{s,j}(T)a_{s,j}AL_jF}+g_{s,j},
\end{align}
where $g_{s,j}$ is a \textcolor{black}{nonlinear} function of $c_{s,j}^{surf}, c_{solv}^{surf}, T_c, I_{cell}$, and $L_{sei}$. 
\textcolor{black}{At the boundary of the particle when $r=R_{s,j}$, the right hand side (RHS) of the boundary condition in \eqref{eq:PDE_solidBound} has a negative sign for the negative electrode, whereas the positive sign is for the positive electrode. The sign is considered to indicate the intercalation and de-intercalation of lithium within the positive and negative electrode.  For instance, when the cell is being discharged ($I_{cell}>0$),  the RHS sign  $\left(\frac {\partial c_{s,n}} {\partial r}\Bigr|_{\substack{r=R_{s,n}}} < 0 \right)$ indicates that lithium is being de-intercalated at the negative electrode (due to the oxidation reaction) and intercalated  $\left( \frac {\partial c_{s,p}} {\partial r}\Bigr|_{\substack{r=R_{s,p}}} > 0 \right)$ at the positive electrode (due to the reduction reaction). During charging ($I_{cell}<0$), the RHS sign indicates intercalation $\left(\frac {\partial c_{s,n}} {\partial r}\Bigr|_{\substack{r=R_{s,n}}} > 0 \right)$ at the negative electrode and de-intercalation at the positive electrode $\left( \frac {\partial c_{s,p}} {\partial r}\Bigr|_{\substack{r=R_{s,p}}} < 0 \right)$.}
 \textcolor{black}{The complete expression of the function $g_{s,j}$ for each electrode is reported in  \eqref{agingfunction}.} We use the FDM to radially discretize the PDEs~\eqref{eq:PDE_solid} into a system of ODEs~\cite{modelT}. 
Solid electrode parameters, including the diffusion coefficient $D_{s,j}$ and the reaction rate
constant $k_{j}$, follow an Arrhenius relationship with
temperature given by 
\begin{align} \label{eq:Arrhenius}
\varphi(T_c) = \varphi_{ref} \exp\left[\dfrac{E_{a,\varphi}}{R_g} \left( \frac{1}{T_{c,ref}}-\frac{1}{T_c}\right)\right]
\end{align}
\textcolor{black}{with $T_{c,ref} = 298K$,  $\varphi$ to be either $D_{s,j}$, $D_{solv}$, or $k_{j}$, and $\varphi_{ref}$ is the value of $\varphi$ at reference temperature $T_{c,ref}$}. \\ 
The surface overpotentials of each electrode, $\eta_j$ for $j\in[n,p]$, \textcolor{black}{are} obtained from the
Butler–Volmer kinetic equation describing the rate of intercalation and de-intercalation of lithium ions as
\begin{equation}
\label{eq:overpotential}
\eta_j = \dfrac{R_g T_c}{0.5F}sinh^{-1}\left(\dfrac{I_{cell}}{2Aa_{s,j}L_ji_{0,j}}\right)	
\quad j\in[n,p]
\end{equation}
where the exchange current density $i_{0,j}$ is given by
\begin{equation}
\label{eq:exchange}
\ i_{0,j} = k_j F \sqrt{c_{e}^{avg}c_{s,j}^{surf}\left(c_{s,j}^{max}-c_{s,j}^{surf}\right)}	
\quad j\in[n,p].
\end{equation}
%
The cell voltage $V_{cell}$ can be calculated as
\begin{align}
\label{eq:cell_voltage}\nonumber
V_{cell} =& U_p(c^{surf}_{s,p}) + \eta_p(c^{surf}_{s,p},c^{avg}_{e},T_c,I_{cell}) - U_n(c^{surf}_{s,n})  \\ 
&- \eta_n (c^{surf}_{s,n},c^{avg}_{e},T_c,I_{cell})- I_{cell}\left(R_{l} + R_{el} + R_{sei}\right)	
\end{align}
in which the cell ohmic resistance includes the 
contact resistance $R_{l}$, electrolyte resistance $R_{el}$, and SEI layer resistance $R_{sei}$, \textcolor{black}{where the last two parameters are expressed as }
\begin{align}
\label{eq:resis}\nonumber
R_{el} &= \frac{1}{2A} \left[\frac{L_n}{\kappa_{e,n}^{eff}} + \frac{2 L_s}{\kappa_{e,s}^{eff}} + \frac{L_p}{\kappa_{e,p}^{eff}} \right], 
 \\
R_{sei} &= \frac{L_{sei}}{a_{s,n} A L_n \kappa_{sei}}, 
\end{align}
where $\kappa_{e,j}^{eff}$ is a function of $c_{e}^{avg}$, and $\epsilon_{e,n}, \epsilon_{e,s}, \epsilon_{e,p}$ are the porosity values in the negative electrode, separator, and positive electrode, respectively. The cell voltage is also dependent on the open circuit potentials of electrodes $U_j$, with $j\in[n,p]$, that are calculated using empirical relationships as functions of electrode surface concentration stoichiometry~\cite{modelT,Allam2019} \textcolor{black}{(also shown in Fig. \ref{fig.OCP} for the cell chemistry used in this study)}.

\textcolor{black}{The bulk SOC of each electrode is given by }
\begin{align} \label{eq:SOC}
SOC^{bulk}_j=\frac{\frac{c^{bulk}_{s,j}}{c^{max}_{s,j}}-\theta^j_{0\%}}{\theta^j_{100\%}-\theta^j_{0\%}}	\qquad j\in[n,p]
\end{align}
that varies between two stoichiometric
values $\theta^j_{100\%}$ and $\theta^j_{0\%}$, representing fully charged
and discharged conditions for each electrode. In this paper, $SOC^{bulk}_p$ is used as battery cell SOC in the optimization algorithm \textcolor{black}{as} the cathode is the limiting electrode.  

\subsection{Cell thermal model}
\textcolor{black}{The thermal dynamics is modeled using the lumped parameter
two-state thermal model }
\begin{align} \label{eq:thermal_core_DE}\nonumber
\ C_c \dfrac{dT_{c}}{dt} &= I_{cell} (V_{oc} - V_{cell}) + \dfrac{T_{s}-T_{c}}{R_c} \\
\ C_s \dfrac{dT_{s}}{dt} &= \dfrac{T_{amb}-T_{s}}{R_u}- \dfrac{T_{s}-T_{c}}{R_c}
\end{align}
where $T_c$ and $T_s$ are the core and surface temperature of each cell. This model assumes that the
internal temperature is uniformly distributed across the core and
the surface temperature is uniform throughout the surface~\cite{Therm-new}.

\subsection{Cell aging model}
A physics-based approach is employed for battery aging that
considers \textcolor{black}{the} anode SEI layer growth as a function of solvent reduction kinetics and diffusion dynamics to predict cell capacity loss and
power fade. \textcolor{black}{For the radial coordinate $r\in[R_{s,n}, R_{s,n}+L_{sei}]$ across the thickness of the SEI layer, the solvent concentration available
for reduction reaction at the anode surface is modeled by}
\begin{align} \label{eq:aging_pde0} 
&\dfrac{\partial c_{solv}}{\partial t} = D_{solv}(\textcolor{black}{T_c}) \dfrac{\partial ^{2} c_{solv}}{\partial r^{2}} - \dfrac{dL_{sei}}{dt} \dfrac{\partial c_{solv}}{\partial r} 
\end{align}
with boundary conditions
\begin{align} \label{eq:aging_pde0Bound} \nonumber
&-D_{solv}(\textcolor{black}{T_c})\dfrac{\partial c_{solv}} {\partial r}\Bigr|_{\substack{r=R_{s,n}}} + \dfrac {dL_{sei}}{dt} c^{surf}_{solv} = \dfrac{i_{s}}{F} \\
&c_{solv}\Bigr|_{\substack{r=R_{s,n}+L_{sei}}} = \epsilon_{sei}c_{solv}^{bulk}.
\end{align}

The PDE aging dynamics~\eqref{eq:aging_pde0} is discretized via FDM where a time-varying grid size is used to account for
changes in the SEI layer thickness~\cite{modelT}.
The SEI layer growth is modeled as follows
\begin{equation} \label{eq:sei_layer_ode}
\dfrac{dL_{sei}}{dt} = -\dfrac{i_{s}M_{sei}}{2F\rho_{sei}}, 
\end{equation}
where the rate of change of $L_{sei}$ is linearly proportional to the side-reaction current 
\begin{align} \label{eq:side_reaction} \nonumber
i_{s} =& -2Fk_{f}(c^{surf}_{s,n})^2c_{solv}^{surf} \\ 
&\exp\left[\dfrac{-\beta F}{R_gT_{c}} \left( \Phi_{s,n} - R_{sei}I_{cell} - U_{s}\right)\right].
\end{align}
\textcolor{black}{The capacity loss is modeled by integrating the side
reaction current as }
\begin{equation} \label{eq:sei_layer_odeQ}
\dfrac{dQ}{dt} = i_{s} {AL_{n}a_{s,n}}.
\end{equation}

\subsection{State-space representation: cell-level}    
Upon discretization, the governing PDEs are transformed into a system of ODEs and DAEs 
using which the cell-level state-space form can be derived. Note that DAEs are related to the Butler-Volmer equation used to calculate the overpotentials.

\textbf{Solid phase diffusion:} the state-space representation of the \textcolor{black}{solid-phase diffusion dynamics for each electrode is represented as}
\begin{align}
&\dot{\boldsymbol{c}}_{s,j} = \alpha_{s,j}A_{s,j}\boldsymbol{c}_{s,j}+\beta_{s,j}B_{s,j}
\left[I_{cell} - g_{s,j}\right], 
\end{align}   
where 
$\boldsymbol{c}_{s,j}=[c_{s,j,1}, \dots, c_{s,j,N_{r,j}}]^T\in \mathbb{R}^{N_{r,j}}$
with $c_{s,j,N_{r,j}} = c_{s,j}^{surf}$,  $B_{s,j} =\left[
0, \dots, (2 + \dfrac{2}{N_r-1})\right]^T\in\mathbb{R}^{N_{r,j}}$,
{\small \begin{align}
	A_{s,j} = \begin{bmatrix}
	-2 & 2 & 0 & 0 & \hdots & 0 & 0 \\
	1/2 & -2 & 3/2 & 0 & \hdots & 0 & 0\\
	0 & 2/3 & -2 & 4/3 & \hdots & 0 & 0\\
	\vdots & \vdots & \vdots & \vdots & \ddots & \vdots & \vdots \\
	0 & 0 & 0 & 0 & \hdots & 2 & -2
	\end{bmatrix}\in\mathbb{R}^{N_{r,j}\times N_{r,j}},
	\end{align}}
\begin{align}
\alpha_{s,j} = \dfrac{D_{s,j}\textcolor{black}{(T_c)}}{\Delta r_j^2} \text{, } \beta_{s,j} = \begin{cases} 
\dfrac{-1}{AL_jFa_{s,j}\Delta r_j} & \text{if } j = n \\
\dfrac{1}{AL_jFa_{s,j}\Delta r_j} & \text{if } j = p \\
\end{cases} \text{, } 
\end{align}
and
{\small\begin{align}
	g_{s,j}(c_{s,j}^{surf},c_{solv}^{surf},T_c, I_{cell} ,L_{sei}) = \begin{cases}
	a_{s,n}L_nAi_s & \text{if } j = n \\
	0 & \text{if } j = p \label{agingfunction}\\ \end{cases} 
	\end{align}}
\noindent with $\Delta r_j=\frac{R_{s,j}}{N_{r,j}-1}$ and $N_{r,j}$ the number of radial discretization grids in SPM.

\textbf{SEI layer growth:} the ODEs for SEI layer growth and capacity loss are given by 
\begin{align} \label{eq:LQ}
\dot{L}_{sei}= \beta_{sei}g_{s,n} \ \ \textrm{and} \ \  \dot{Q}=\frac{\dot{L}_{sei}}{\beta_{sei}}= a_{s,n}L_nAi_s 
\end{align}
with $\beta_{sei} = \dfrac{-M_{sei}}{2F\rho_{sei}a_{s,n}L_nA}$.

\textbf{Solvent diffusion:} 
\textcolor{black}{the ODEs describing the solvent diffusion dynamics is given by}

{\small \begin{align}
	\label{eq:discrete_solv_dynamics}
	\dot{\boldsymbol{c}}_{solv} &= \begin{cases}
	&2\alpha_{solv}(c_{solv,2} - c_{solv,1})+ \\ &\qquad \beta_{solv}\left(\dfrac{i_s}{F} - \dfrac{dL_{sei}}{dt}c_{solv,1}\right),\ \text{ if } i = 1 \\
	&\alpha_{solv}\left(c_{solv,i+1}-2c_{solv,i}+c_{solv,i-1}\right)+ \\ &\qquad \gamma_{solv}\left(c_{solv,i+1}-c_{solv,i-1}\right),\ \text{ if 1$<$i$<$$N_{sei}$} \\
	&0,\ \text{ if } i = N_{sei} \\
	\end{cases}  
	\end{align}}	
	
with
$ \alpha_{solv}=\dfrac{D_{solv}(\textcolor{black}{T_c})}{\left(L_{sei}\Delta\xi\right)^2} $, 
$\gamma_{solv}=\left(\dfrac{\xi-1}{2L_{sei}\Delta\xi}\dfrac{dL_{sei}}{dt}\right) $ and
$\beta_{solv}=\left(\dfrac{2}{L_{sei}\Delta\xi}+\dfrac{1}{D_{solv}(\textcolor{black}{T_c})}\dfrac{dL_{sei}}{dt}\right)$.

\noindent where $\boldsymbol{c}_{solv}=[c_{solv,1},\dots,c_{solv,N_{sei}}]^T\in \mathbb{R}^{N_{sei}}$ with $c_{solv,1} = c_{solv}^{surf}$; and $\xi=\frac{r-R_{s,n}}{L_{sei}}$ and $\Delta\xi=\frac{1}{N_{sei}-1}$ with $N_{sei}$ as the number of \textcolor{black}{SEI layer discretization points \cite{TreyModel}}.

\subsection{Surrogate model for solvent diffusion dynamics} 

In the cell model, the aging dynamics, inclusive of the SEI layer growth and solvent diffusion acts as the slow dynamics. 
In particular, the characteristic time scales of the battery dynamics can be calculated as~\cite{Allam2019} 
\begin{equation} \label{eq.time}
t_{ter}=\frac{R^2_{cell}}{\phi}, \ \ t_{elec}=\frac{R^2_{s,n}}{D_{s,n}}, \ \ t_{ag}=\frac{R^2_{s,n}}{D_{solv}}, 
\end{equation} 
where $t_{ter}$, $t_{elec}$, and $t_{ag}$ are the time scales of the thermal, electrochemical, and aging dynamics, respectively, \textcolor{black}{$R_{cell}$ is the radius of a cylindrical lithium-ion cell, $\phi$ is the thermal diffusivity,  $R_{s,n}$ is the particle radius in the negative electrode, $D_{s,n}$ is the solid-phase diffusion in the negative electrode, and $D_{solv}$ is the solvent diffusion.} Incorporating parameter values from the literature~\cite{TER1,TER2,Allam2019} shows that $t_{ag}$ is in the order of $10^8$ $sec$ while $t_{ter}$ and $t_{elec}$ are in the orders of $10-100$ $sec$ and $10^3$ $sec$, respectively, implying that the cell model is a three-time scale system in which $t_{ter}< t_{elec}\ll t_{ag}$.

The difference in temporal scales in the cell dynamics is the cause of long - at time, prohibitive - simulation times which are not compatible with the design of an optimization strategy. 
In the aging dynamics, the SEI layer growth is the low-dimensional slow variable whose dimension is determined by the number of cells in the battery modules, whereas the solvent concentration dynamics is a high-dimensional state whose dimension is dependent on the number of discretization points of the solvent diffusion PDE.

\textcolor{black}{The integration of solvent diffusion dynamics~\eqref{eq:aging_pde0} represents the  major bottle neck from a computational standpoint.}
To get a fast simulation time, we propose a surrogate model to  capture the solvent diffusion dynamics~\eqref{eq:discrete_solv_dynamics} based on a  joint optimization/curve fitting approach  (see Fig.~\ref{fig.joint}). 
The surrogate model is built to identify a constant value of $c_{solv}$ as a function of $I_{cell}$  and $T_{amb}$ to ensure that the final value of the SEI layer thickness from the high fidelity model is accurately predicted.
Note that the solvent concentration $c_{solv}^{surf}$ is used to calculate the side-reaction current~\eqref{eq:side_reaction} based on which SEI layer growth and cell capacity loss are calculated (see ~\eqref{eq:sei_layer_ode}). 

\begin{figure*}[t]
	\centering
		\includegraphics[scale=0.48]{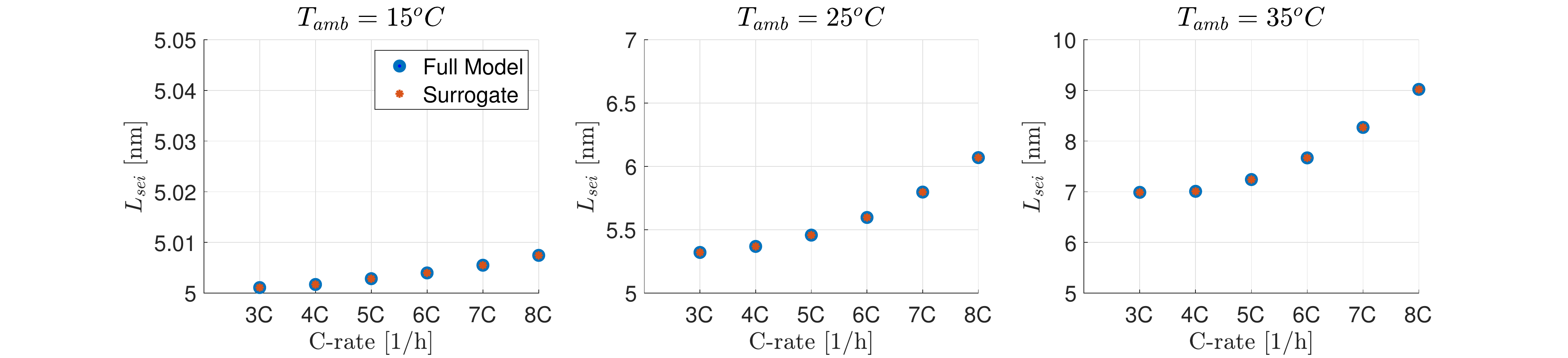}
	\caption{\textcolor{black}{Comparison of the SEI layer thickness values between the SPM with solvent-diffusion dynamics and the proposed surrogate model at the end of charging cycles at C-rates $ =  [3C, 4C, 5C, 6C, 7C, 8C]$ at three different ambient temperatures $T_{amb} = [15^{o}C, 25^{o}C, 35^{o}C]$.}}
	\label{fig:Surrogate_Comp}
\end{figure*}

\textcolor{black}{The following unconstrained optimization problem is formulated to find the optimal $c_{solv}^{surf*}$ 
\begin{equation} \label{eq.Csolve}
c^{surf*}_{solv}=
\underset{c_{solv}}{\textrm{min}} \|L^{hf}_{sei}-L^{lf}_{sei}(c^{surf}_{solv})\|,
\end{equation} 
where $L^{hf}_{sei}$ is the SEI layer thickness from the SPM inclusive of the solvent diffusion model ~\eqref{eq:discrete_solv_dynamics}, whereas $L^{lf}_{sei}({c^{surf}_{solv}})$ is the SEI layer thickness when  constant solvent 
\textcolor{black}{concentration} 
is used. }
\textcolor{black}{Note that the SEI layer thickness values $L^{hf}_{sei}$  and $L^{lf}_{sei}({c^{surf}_{solv}})$ are the final values at the end of the charging time. In Figure \ref{fig:Surrogate_Comp}, the difference between the final SEI layer thickness values from the SPM with solvent-diffusion dynamics $L^{hf}_{sei}$ and the SEI layer thickness from the surrogate model $L^{lf}_{sei}({c^{surf*}_{solv}})$ is shown for six different charging C-rates of $[3C, 4C, 5C, 6C, 7C, 8C]$ at three different ambient temperatures $T_{amb} = [15^{o}C, 25^{o}C, 35^{o}C]$. As observed,  the SEI layer thickness values coincide with each other, thereby proving that the surrogate model is a suitable choice to replace the higher dimensional model to solve the optimal control problem successfully with lower computation cost. The resulting optimal values of $c^{surf*}_{solv}$ obtained from the unconstrained optimization problem are fitted as a function of $I_{cell}$ and $T_{amb}$ 
using $5^{th}$-order polynomials.}



\textcolor{black}{\begin{center}
\begin{figure}\centering
\hspace*{-3em}
	\includegraphics[scale=0.4]{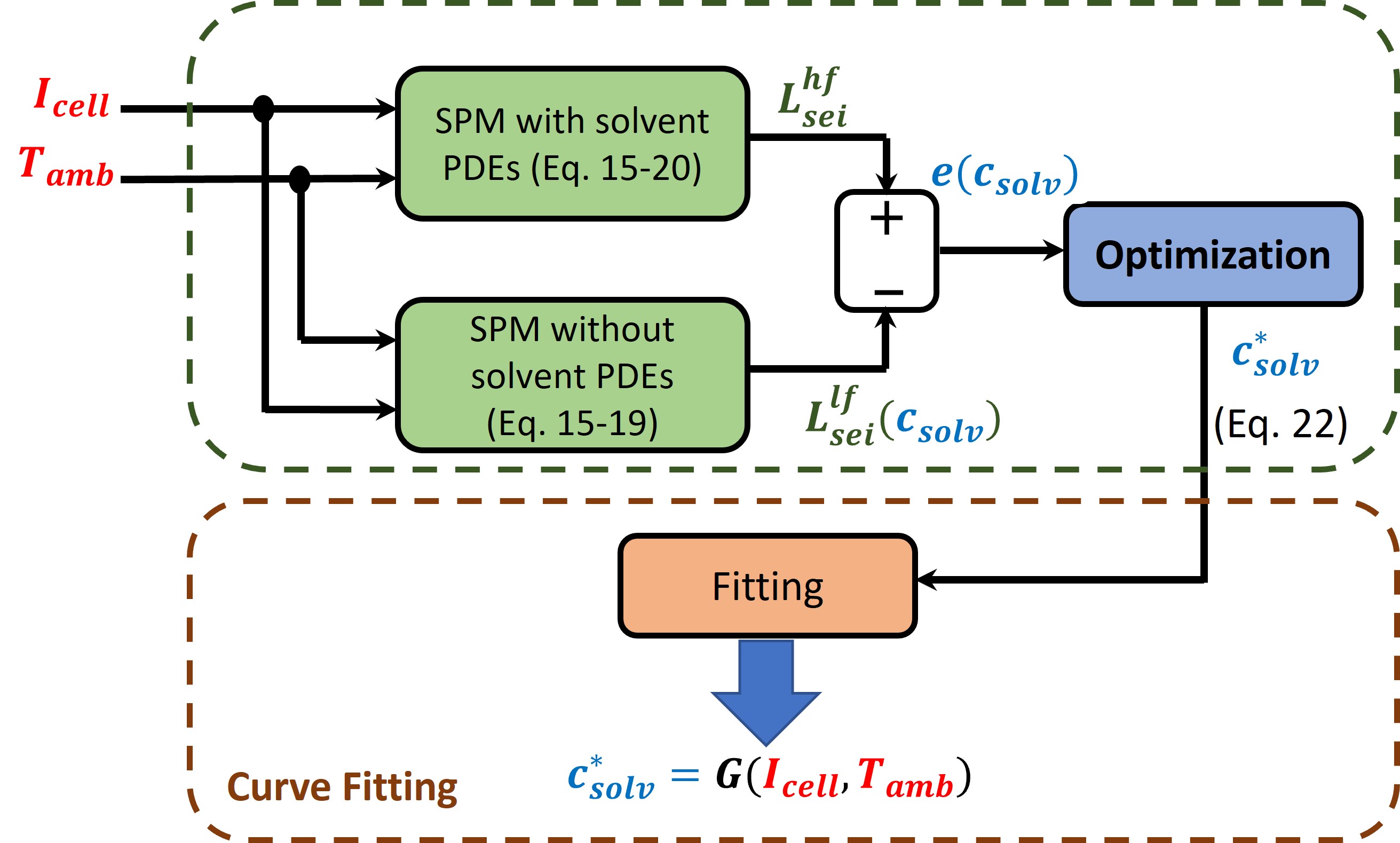}
	\vspace{0.5em}
	\caption{Scheme of the surrogate model derivation to calculate $c_{solv}^{*}$.}
	\label{fig.joint}
\end{figure} 
\end{center}}

{\begin{table*}[ht]
		\caption{Module-level matrices and state vectors.}
		\small
		\label{table:block_diag}
		\textbf{Solid phase diffusion:}
		$A_{s,j}^{mod} = \begin{bmatrix}
		\left(\alpha_{s,j}A_{s,j}\right)_{1} & & \\
		& \ddots & \\
		& & \left(\alpha_{s,j}A_{s,j}\right)_{N_{cell}}
		\end{bmatrix}$, \ \ \
		$B_{s,j}^{mod} = \begin{bmatrix}
		\left(\beta_{s,j} B_{s,j}\right)_{1} \\
		\vdots \\
		\left(\beta_{s,j} B_{s,j}\right)_{N_{cell}} 
		\end{bmatrix}$, \ \ \
		$\boldsymbol{G}_{s,j}^{mod} = \begin{bmatrix}
		\left[\beta_{s,j}B_{s,j}g_{s,j}\right]_{1}\\
		\vdots \\
		\left[\beta_{s,j}B_{s,j}g_{s,j}\right]_{N_{cell}}
		\end{bmatrix}$
		\\
		\textbf{Thermal:}
		$A_{therm}^{mod} = \begin{bmatrix}
		\dfrac{-1}{R_cC_c} & \dfrac{1}{R_cC_c} & \hdots &  0 \\
		\dfrac{1}{R_cC_s} & \dfrac{-1}{C_s}\left(\dfrac{1}{R_c}+\dfrac{1}{R_u}+\dfrac{1}{R_m}\right) & \hdots & 0\\
		0 & 0 & \hdots &  0\\
		0 & \dfrac{1}{R_mC_s} & \hdots & 0\\
		\vdots & \vdots & \ddots & \vdots\\
		0 & 0 & \hdots &  \dfrac{-1}{C_s}\left(\dfrac{1}{R_c}+\dfrac{1}{R_u}+\dfrac{1}{R_m}\right)
		\end{bmatrix}$,
		\\
		$B_{therm}^{mod} = \begin{bmatrix}
		\dfrac{1}{C_c}\left(V_{oc}-V_{cell}\right)_{1}\\
		0 \\
		\dfrac{1}{C_c}\left(V_{oc}-V_{cell}\right)_{2} \\
		0 \\
		\vdots \\
		0 \\
		\end{bmatrix}$, \ \ \ 
		$\boldsymbol{G}_{therm}^{mod} = \begin{bmatrix}
		0 \\
		\dfrac{1}{R_uC_s}\\
		0 \\
		\dfrac{1}{R_uC_s}\\
		\vdots \\
		\dfrac{1}{R_uC_s}\\
		\end{bmatrix}$ 
		\\ \textbf{Aging:}
		$\boldsymbol{G}_{sei}^{mod} = \begin{bmatrix}
		\left[\beta_{sei}g_{s,n}\right]_{1}\\
		\vdots \\
		\left[\beta_{sei}g_{s,n}\right]_{N_{cell}}
		\end{bmatrix}$, \ \ \ \ \ \ \
		$\boldsymbol{G}_{Q}^{mod} = \begin{bmatrix}
		\left[g_{s,n}\right]_{1}\\
		\vdots \\
		\left[g_{s,n}\right]_{N_{cell}}
		\end{bmatrix}$
		\\ \textbf{State vectors:}
		$\boldsymbol{c}_{s,j}^{mod}= \begin{bmatrix}
		\boldsymbol{c}_{s,j_1}\\
		\vdots\\
		\boldsymbol{c}_{s,j_{N_{cell}}}
		\end{bmatrix}$, \ \ \
		$\boldsymbol{T}^{mod}=\begin{bmatrix}
		T_{c_1} \\ T_{s_1} \\ \vdots \\ T_{c_{N_{cell}}} \\ T_{s_{N_{cell}}}
		\end{bmatrix}$, \ \ \
		$\boldsymbol{L}_{sei}^{mod}= \begin{bmatrix}
		L_{sei_1}\\
		\vdots\\
		L_{sei_{N_{cell}}}
		\end{bmatrix}$, \ \ \ 
		$\boldsymbol{Q}^{mod}= \begin{bmatrix}
		Q_{1}\\
		\vdots\\
		Q_{N_{cell}}
		\end{bmatrix}$
		\vspace*{-2em}
\end{table*}}

\vspace{-1.25cm}
\subsection{State-space representation: module-level}  
\textcolor{black}{The state and parameter heterogeneity due to manufacturing imperfections and \textcolor{black}{non-uniform} operating conditions \textcolor{black}{can cause} exacerbated aging of \textcolor{black}{the} battery pack when compared to a \textcolor{black}{single} cell.
The overall thermal and aging effects of the cells \textcolor{black}{in a module} can be captured \textcolor{black}{through} heat transfer between cells.
\textcolor{black}{The thermal interconnection between adjacent cells in the battery module is provided via the thermal resistance term $R_m$ among a cell $k$ with the downstream and upstream cells,  $k-1$ and $k+1$, respectively.} \textcolor{black}{This results in surface temperature dynamics of interconnected cells that are modeled} 
as follows ~\cite{MULTI1}}
\begin{align} \label{eq:thermal_surf_DE_pack}
\hspace{-15mm}
C_s \dfrac{dT_{s_k}}{dt} &= \dfrac{T_{amb}-T_{s_k}}{R_u}
- \dfrac{T_{s_k}-T_{c_k}}{R_c} + \nonumber \\ &\dfrac{T_{s_k}-T_{s_{k+1}}}{R_m} +
\dfrac{T_{s_k}-T_{s_{k-1}}}{R_m}.
\end{align}
The core temperature of cell $k$ is resolved using the relation already stated in~\eqref{eq:thermal_core_DE}.
In the  module-level matrix $A_{therm}^{mod}$ in Table~\ref{table:block_diag} the surface temperature states, $T_{s_k}$,  embed the cell-to-cell heat transfer from \eqref{eq:thermal_surf_DE_pack}.

A convenient shorthand \textcolor{black}{term} 
for module-level dynamics with $N_{cell}$ series-connected cells is as follows 
\begin{align}
\label{eq:pack_dynamicsMain} \nonumber
\dot{\boldsymbol{c}}_{s,j}^{mod} &= A_{s,j}^{mod}\boldsymbol{c}_{s,j}^{mod}+ B_{s,j}^{mod}\boldsymbol{u} - \boldsymbol{G}_{s,j}^{mod}\\ \nonumber
\dot{\boldsymbol{T}}^{mod} &= A_{therm}^{mod}\boldsymbol{T}^{mod}+B_{therm}^{mod}\boldsymbol{u} + \boldsymbol{G}_{therm}^{mod}T_{amb} \\ \nonumber
\dot{\boldsymbol{L}}_{sei}^{mod} &= \boldsymbol{G}_{sei}^{mod} \\ \nonumber
\dot{\boldsymbol{Q}}^{mod}&= \boldsymbol{G}_{Q}^{mod}\\ 
\dot{\boldsymbol{c}}_{solv}^{mod}&= \boldsymbol{G}_{solv}^{mod}
\end{align}
where $\boldsymbol{u}$ includes the currents of all cells, the module-level block diagonal coefficient matrices and state vectors are listed in Table~\ref{table:block_diag}, and the module state vector at the \textit{system level} is 
\begin{align}
\boldsymbol{z}(t)=[\boldsymbol{c}_{s,j}^{mod} \; \boldsymbol{T}^{mod} \;
\boldsymbol{L}_{sei}^{mod} \; \boldsymbol{Q}^{mod} \; \boldsymbol{c}_{solv}]^T.
\end{align}

\textcolor{black}{Note that the right hand side of the solvent diffusion dynamics,  $\bold{G}_{solv}^{mod}$, is a
nonlinear function of the states and input that can be derived for each cell using~\eqref{eq:discrete_solv_dynamics}.}
It should be also pointed out that $\alpha_{s,j}$ and $\beta_{s,j}$ used in $A_{s,j}^{mod}$, $B_{s,j}^{mod}$, and $\boldsymbol{G}_{s,j}^{mod}$ vary between cells due to cell heterogeneity
of design parameters, non-uniform aging,
temperature distribution.

Upon the creation of the surrogate model to replace the  solvent diffusion dynamics, presented in the previous subsection, the module state vector used in the proposed optimal control design is given by~\footnote{\textcolor{black}{Note that  $\boldsymbol{c}_{solv}$ included in $\boldsymbol{z}(t)$ is now excluded from the \textit{system-level} state vector $\boldsymbol{x}(t)$ due to the inclusion of the surrogate model.}}
\begin{align}\label{eq:sst}
\boldsymbol{x}(t)=[\boldsymbol{c}_{s,j}^{mod} \;\boldsymbol{T}^{mod} \;
\boldsymbol{L}_{sei}^{mod} \; \boldsymbol{Q}^{mod}]^T \in\Re^{N_{s}},
\end{align}
where the number of states is $N_{s}=N_{cell}\left(4+2(N_r-1)\right)$ for a given $N_r$.
\section{Optimal Control Problem Formulation} \label{section.ps}

\begin{figure*}\centering
	\includegraphics[scale=0.45]{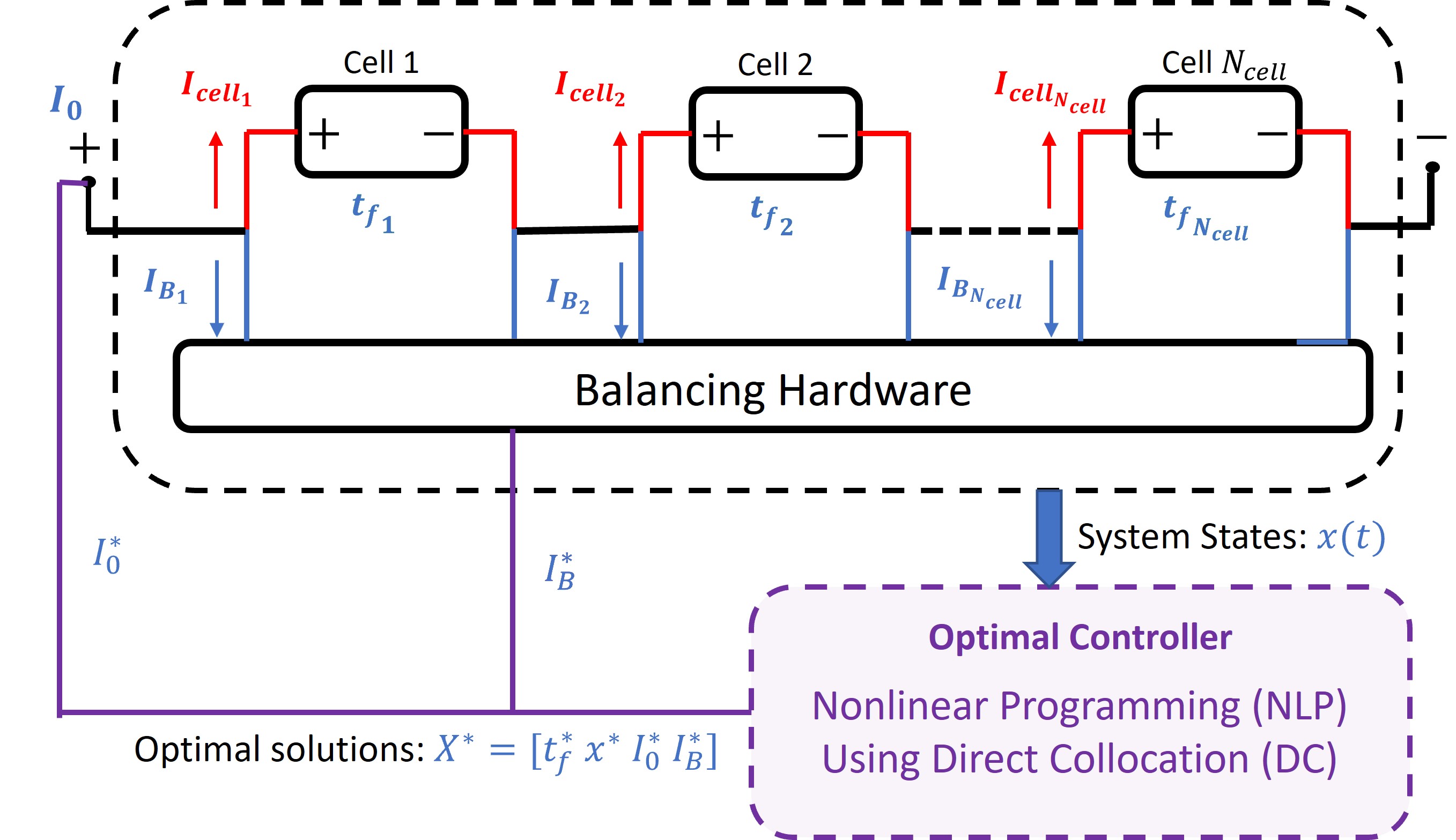}
	\caption{\textcolor{black}{Battery module with $N_{cell}$ series-connected cells, where each cell is connected to a balancing hardware. For both schemes (OCP-SCT and OCP-DCT), the proposed  optimal controller realized through Nonlinear Programming using Direct Collocation optimizes the variables annotated in blue for fast charging and minimized degradation.}}
	\label{fig.Pack5}
\end{figure*}

\textcolor{black}{
In this section, we formulate a multi-objective optimal control framework for fast charging and minimum degradation of a battery module with $N_{cell}$ series-connected imbalanced cells as shown in Fig.~\ref{fig.Pack5}.} In this configuration, 
the module capacity \textcolor{black}{is limited by the}
\textcolor{black}{capacity of the} weakest cell in the string.
%
%
%
%
\textcolor{black}{In this work, we account for the intrinsic heterogeneity among the cells in terms of charge, temperature and  $SOH$. Moreover, battery health is defined both in terms of $Q$ and $R_{sei}$,   both dependent on $L_{sei}$ as seen from ~\eqref{eq:resis} and~\eqref{eq:LQ}.
	To model cells \textcolor{black}{subject to} $SOH$ imbalances, selection of different initial conditions for $L_{sei}$ is made.}


A \textcolor{black}{multi-objectve} OCP is formulated for two different \textcolor{black}{charging schemes,} OCP-SCT and OCP-DCT. The former assumes that all cells are charged simultaneously, irrespective of their non homogeneous initial states,  whereas the latter assumes different times of charging of  the cell to reflect the non uniform initial states the cells are at. In particular, OCP-DCT is aimed at providing a \textcolor{black}{charging} strategy that extends the battery life and provides more flexibility against heterogeneity among the cells.
From Fig.~\ref{fig.Pack5}, $I_0=I_{cell_k}+I_{B_k}$ for $k=1,\dots,N_{cell}$ from which one can define the vector
\begin{equation} \label{eq.Icell}
\boldsymbol{I_{cell}}=[I_{cell_{1}} \dots I_{cell_{N_{cell}}}]^T=
[I_0-I_{B_1}\dots I_0-I_{B_{N_{cell}}}]^T.
\end{equation}
During charging, the module current $I_0\in\Re$ and the vector of balancing currents  $\boldsymbol{I_B}=[I_{B_1} \dots I_{B_{N_{cell}}}]^T\in\Re^{N_{cell}}$ are unknown and optimally planned.
 \textcolor{black}{Each cell $k$ is connected in parallel to an active balancing circuitry whose current $I_{B_{k}}$ is determined by the proposed optimal controller.}
The following OCP for the OCP-DCT scheme is formulated as:
\begin{align} \label{eq.OCP}
\boldsymbol{X}^*=
\underset{\boldsymbol{X}\in\Re^{N_{opt}}}{\textrm{argmin}} &\alpha\beta_1h(\boldsymbol{t_f})+  \\
&(1-\alpha) \left(\beta_2g_1(\boldsymbol{L_{sei}})+\beta_3g_2(\boldsymbol{\dot{L}_{sei}})\right)\nonumber,
\end{align}  
where the vector of optimization variables $\boldsymbol{X}$ is comprised of the vector of final times of charging $\boldsymbol{t_f}=[t_{f_1} \dots t_{f_{{N_{cell}}}}]^T\in\Re^{N_{cell}}$, the system state $\boldsymbol{x}(t)\in\Re^{N_{s}}$, the module current $I_0$, and the balancing current vector $\boldsymbol{I_B}$:  
\begin{equation} \label{eq.optvar}
\boldsymbol{X}=\left[\boldsymbol{t_f},\boldsymbol{x}(t),I_0(t),\boldsymbol{I_B}(t)\right]^T\in\Re^{N_{opt}}.
\end{equation}
The number of optimization variables is $N_{opt}=N_{s}+2N_{cell}+1$ and the continuously differentiable functions $g_1$, $g_2$, and $h$ are defined as
\begin{align} \label{eq.g1g2} \nonumber
g_1(\boldsymbol{L_{sei}})=&\frac{1}{N_{cell}}\sum_{k=1}^{N_{cell}}L_{sei_k},
\\   \nonumber
g_2(\boldsymbol{\dot{L}_{sei}})&=\frac{1}{N_{cell}}\sum_{k=1}^{N_{cell}}\dot{L}_{sei_k},
\\ 
h(\boldsymbol{t_f})=&\frac{1}{N_{cell}}\sum_{k=1}^{N_{cell}}t_{f_k}.  
\end{align}

Note that $g_1(\boldsymbol{L_{sei}}), g_2(\boldsymbol{\dot{L}_{sei}})$, and $h(\boldsymbol{t_f})$ are operators that return the average of SEI layer thicknesses \textcolor{black}{at the end of charging}, 
\textcolor{black}{the average of the SEI layer thickness growth rates}, 
and the average of charging times, respectively. Thus, the OCP~\eqref{eq.OCP} along with the definitions~\eqref{eq.g1g2} forms a \textit{min-mean optimization problem}. 
\textcolor{black}{The positive scalars $\beta_1$ $[\text{s}^{-1}]$, $\beta_2$ $[\text{sm}^{-1}]$, and $\beta_3$ $[\text{sm}^{-1}]$ are optimization weights corresponding to the charging time and SEI layer growth objectives, respectively, which are chosen prior to our exploration of the \textcolor{black}{parameter space to} set the objective terms on the same order of magnitude. The dimensionless scalar $0\leq\alpha\leq1$ is a trade-off coefficient that can be adjusted to give three different paradigms: fast charging ($\alpha=1$), minimum degradation ($\alpha=0$), and balanced charging-degradation ($0<\alpha<1$), as demonstrated in Section \ref{subsec.pareto}}.    

The operation of the battery module is subject to the dynamic constraints~\eqref{eq:pack_dynamicsMain} and the following operating constraints for each cell with $k=1,\dots,N_{cell}$.  To establish safety metrics, module and balancing currents, voltages, core and surface temperatures, and solid concentrations of all cells are enforced to lie within \textcolor{black}{their} physical bounds	for $k=1,\dots,N_{cell}$
\begin{align} \label{eq.const-I} \nonumber
I_{B_{\textrm{min}}} &\leq I_{B_k}(t) \leq I_{B_{\textrm{max}}}, \ \  
I_{0_{\textrm{min}}} \leq I_{0}(t) \leq I_{0_{\textrm{max}}} \\ \nonumber
V_{cell{\textrm{min}}} &\leq V_{cell_k}(t) \leq V_{cell{\textrm{max}}} \\ 
T_{lk_{\textrm{min}}}&\leq T_{lk}(t) \leq T_{lk_{\textrm{max}}}, \ \ l\in\{c,s\} \\ \nonumber
\theta^j_{0\%} c_{s,jk_{\textrm{max}}} &\leq c_{s_jk}(t) \leq \theta^j_{100\%}c_{s,jk_{\textrm{max}}}, \ \ j\in\{n,p\}.	
\end{align}	

Initial conditions of the states are taken into consideration as equality constraints 
\begin{align} \label{eq.const-init}  \nonumber
L_{sei_k}(t_0)&=L_{sei_{0_k}}, \ \ Q_k(t_0)=Q_{0_k} \\ 
T_{lk}(t_0)&=T_{lk_0}, \ \ l\in\{c,s\} \\ SOC_k(t_0) &= SOC_{\textrm{initial}_k}
\end{align}
and cells are charged to the same targeted SOC 
\begin{align} \label{eq.const-soc2} 
SOC_k(t_{f_k}) = SOC_{\textrm{target}}.
\end{align}
In the OCP-DCT scheme, charging time is allowed to be different for each cell and an upper bound on the  charging time is imposed as well:
\begin{align} \label{eq.const-t} 
0 \leq t_{f_k}\leq t_{f_{\textrm{max}}}.
\end{align}

A second  problem in the context of optimal charging is investigated in this work  where simultaneous charging time  of all cells in the module must be achieved. We refer to this formulation as the OCP-SCT.

The OCP-SCT problem resembles the OCP-DCT scheme except for the following differences
\begin{enumerate}
	\item The final times of charging are the same for all cells from which one can consider $t_f\in\Re$.
	\item The number of optimization variables reduces to $N_{opt}=N_{s}+N_{cell}+2$; hence
	\begin{equation} \label{eq.optvar-reduc}
	\boldsymbol{X}=\left[t_f,\boldsymbol{x}(t),I_0(t),\boldsymbol{I_B}(t)\right]^T\in\Re^{N_{opt}},
	\end{equation}
	in which $t_f$ is now a scalar. 
	\item The cost function associated with charging in \eqref{eq.OCP} reduces to $h(t_f)=t_f$. 
	\item The constraint associated with the charging time reduces to $0 \leq t_{f}\leq t_{f_{\textrm{max}}}$ in which the charging time of all cells is the same. 
\end{enumerate}
In Section \ref{section.cont}, we solve the OCP-SCT and OCP-DCT subject to dynamic constraints~\eqref{eq:pack_dynamicsMain} and the operating constraints~\eqref{eq.const-I}-~\eqref{eq.const-t}.  

\textcolor{black}{
	\begin{remark}
	Note that the vector of optimization variables $\boldsymbol{X}$ proposed in \eqref{eq.optvar} and \eqref{eq.optvar-reduc}	considers $I_0$ to be a free variable. Only a portion of the module current $I_0$ provided by the optimal solution flows through the cells ($\boldsymbol{I_{cell}}$), meaning that some of the $I_0$ is wasted (by bleeding through the balancing hardware). However, since the value of $I_0$ does not directly affect the objective functions in \eqref{eq.OCP}, the resulting optimal solution in terms of fast charging while ensuring minimum degradation is guaranteed. The rationale behind having $I_0$ and $\boldsymbol{I_B}$ in the optimization variable $\boldsymbol{X}$ is to build on it in the future work by including additional objective functions such as 
		\begin{enumerate}
			\item minimization of module energy consumption (which is dependent on $I_0$), and
			\item minimization of the temperature or heat loss in any general balancing hardware (which is dependent on $\boldsymbol{I_B}$).
		\end{enumerate}  
		In the proposed formulation, since the objective functions are not explicitly minimizing the energy consumption, the value of $I_0$ is not the optimal controller's immediate concern. To that end, there exists an alternate formulation of the optimization variable vector $\boldsymbol{X}$ that can potentially be used to solve the same optimal control problem by eliminating $I_0$ and $\boldsymbol{I_B}$ from the vector of optimization variables and replace it with the cell currents $\boldsymbol{I_{cell}}$, given by
	\begin{equation}\label{eq:new}
	\boldsymbol{X}=\left[\boldsymbol{t_f},\boldsymbol{x}(t),\boldsymbol{I_{cell}}(t)\right]^T,
	\end{equation}
	that satisfy the following algebraic relationships 
	\begin{align}
	I_{0}&=\min \left(I_{cell_{1}} \hdots I_{cell_{N_{cell}}} \right), \nonumber \\
	\boldsymbol{I_{B}}&= \left[I_0 - I_{cell_{1}} \hdots I_{0} - I_{cel_{N_{cell}}} \right].
\end{align}	
	This formulation results in one less optimization variable (without the $I_0$ variable), however, it may not be preferred in the future when objective functions penalizing module energy consumption and temperature or heat loss in the balancing hardware are to be incorporated.
\end{remark}
}

\section{Optimal Control Algorithm} \label{section.cont}


In this paper,  the direct collocation method~\cite{DC1} is employed to solve the OCP characterized by nonlinear coupled dynamic constraints~\eqref{eq:pack_dynamicsMain}.
The original OCP~\eqref{eq.OCP} is transcribed into a NLP problem~\cite{NLP} by approximating all elements of the unknown vector $\boldsymbol{X}$ with  polynomial splines.  Spline approximation refers to the operation of replacing a continuous trajectory with  a sequence of polynomial segments that are glued together at given break points (BPs). 

This results in all trajectories to be discretized in time $0=t_0<t_1<\dots<t_{N_{BP}}=t_f$, where $N_{BP}$ is the number of BPs, and $t_0$ and $t_f$ are the initial and final times, respectively. The order of polynomial segments, $d$, and the degree of smoothness over the BPs, $s$, are specified in such a way that the continuity of discretized trajectories at BPs and between them is ensured. A spline can be parameterized as the weighted sum of B-splines---piecewise polynomials of order $d$---such that each optimization variable vector can be approximated as
\begin{equation}\label{eq.BSpline}
X_p(t)=\sum_{q=1}^{N_{FP_p}}\mathcal{B}_{p,q}\omega_{p,q} \ \ \textrm{for} \ \ p=1,\dots,N_{opt},
\end{equation} 
where $\mathcal{B}_{p,q}$ and $\omega_{p,q}$ are the $q^{th}$ B-spline and \textcolor{black}{free parameters} of the $p^{th}$ optimization variable, and $N_{FP_p}=N_{P}(d_p-s_p)+s_p$ is the number of free parameters for the $p^{th}$ optimization variable with $N_{P}=N_{BP}-1$ as the number of polynomial segments~\cite{Optra}. By parameterizing all of the system trajectories $\boldsymbol{t_f}$, $\boldsymbol{x}(t)$, $I_0(t)$, and $\boldsymbol{I_B}(t)$ ($t_f$ is scalar in case of OCP-SCT), the total number of free parameters \textcolor{black}{are} calculated as
\begin{equation}\label{eq.NFP}
N^{t}_{FP}=N_{FP_x}N_s+(N_{FP_{I_B}}+N_{FP_{t_f}})N_{cell}+N_{FP_{I_0}},
\end{equation}             
where $N_{FP_x}$, $N_{FP_{I_B}}$, $N_{FP_{I_0}}$, and $N_{FP_{t_f}}$ are the numbers of free parameters for each state, balancing and module currents, and charging times, respectively. These are design parameters to be selected by users.

With this approximation in hand, the original OCP~\eqref{eq.OCP}
\textcolor{black}{is} transcribed to the NLP problem as follows
\begin{align}\label{eq.NLP}
\mathbf{P}^*=&
\underset{\boldsymbol{P}}{\textrm{argmin}} \ \ J(\boldsymbol{P}) \\ \nonumber
&\textrm{s.t.} \\ \nonumber
& \boldsymbol{g_{P_1}(P)=0}, \ \ \boldsymbol{g_{P_2}(P)\leq 0}, \ \ \boldsymbol{P_{\textrm{min}}}\leq \boldsymbol{P} \leq \boldsymbol{P_{\textrm{max}}},
\end{align}
where $\boldsymbol{P}=[\omega_{p,q}]\in \Re^{N^{t}_{FP}}$ is the finite set of free parameters; and $J\in\Re$, and $\boldsymbol{g_{P_1}}\in\Re^{m_1}$ and $\boldsymbol{g_{P_2}}\in\Re^{m_2}$
are the cost, and the vectors of linear/nonlinear equality and inequality constraints, respectively, 
all expressed in terms of the vector of the static parameters $\boldsymbol{P}$.    

The Lagrangian function $\mathcal{L}:\Re^{N^{t}_{FP}}\times\Re^{m_1}\times\Re^{m_2}\rightarrow\Re$ associated with the NLP problem~\eqref{eq.NLP} is defined as
\begin{align}\label{eq.lag}
\mathcal{L}(\boldsymbol{P},\boldsymbol{\mu}_1,\boldsymbol{\mu}_2)=J(\boldsymbol{P})+\boldsymbol{\mu}_1^T\boldsymbol{g_{P_1}}(\boldsymbol{P})+\boldsymbol{\mu}_2^T\boldsymbol{g_{P_2}}(\boldsymbol{P})
\end{align}
with $\boldsymbol{\mu}_1\in\Re^{m_1}$ and $\boldsymbol{\mu}_2\in\Re^{m_2}$. The Karush-Kuhn-Tucker (KKT) optimality conditions~\cite{BOYD} associated with~\eqref{eq.lag} are
\begin{align}\label{eq.kkt1} \nonumber
&\nabla\mathcal{L}=\nabla J(\boldsymbol{P}^*)+\sum_{r=1}^{m_1}\mu_{1_r}^*\nabla g_{P_{1_r}}(\boldsymbol{P}^*) \\
&+\sum_{r=1}^{m_2}\mu_{2_r}^*\nabla g_{P_{2_r}}(\boldsymbol{P}^*)=0 \ \ (\textrm{Stationarity}), 
\\ \label{eq.kkt2} \nonumber
& \boldsymbol{g_{P_1}}(\boldsymbol{P}^*)=0 \ \ \textrm{for} \ \ r=1,\dots,m_1 \\   
&\boldsymbol{g_{P_2}}(\boldsymbol{P}^*)\leq0 \ \ \textrm{for} \ \ r=1,\dots,m_2 \ \ (\textrm{Primal feasibility}),
\\ \label{eq.kkt3}
&\mu_{2_r}^*\geq 0 \ \ \textrm{for} \ \ r=1,\dots,m_2 \ \ (\textrm{Dual feasibility}),
\\ \label{eq.kkt4}
&\sum_{r=1}^{m_2}\mu_{2_r}^*g_{P_{2_r}}(\boldsymbol{P}^*)=0 \ \ (\textrm{Complementary slackness}),
\end{align}

\noindent where conditions~\eqref{eq.kkt1}-\eqref{eq.kkt4} are called \textit{Stationarity}, \textit{Primal feasibility}, \textit{Dual feasibility}, and \textit{Complementary slackness}, respectively, and $\mu_{1_r}$ for $r=1,\dots,m_1$ and $\mu_{2_r}$ for $r=1,\dots,m_2$ are KKT multipliers. For any continuously differentiable cost $J$ and constraints $\boldsymbol{g_{P_1}}$ and $\boldsymbol{g_{P_2}}$, if there exists a pair of $(\boldsymbol{\mu}_{1}^*,\boldsymbol{\mu}_{2}^*)$ such that the KKT conditions~\eqref{eq.kkt1}-\eqref{eq.kkt4} hold, then a solution $\boldsymbol{P}^*$ is a local optimum for the NLP problem~\eqref{eq.NLP}. It should be pointed out that when $\boldsymbol{P}^*$ and $(\boldsymbol{\mu}_{1}^*,\boldsymbol{\mu}_{2}^*)$ are any primal dual optimal points with zero duality gap (strong duality), then any pair of $(\boldsymbol{P}^*,(\boldsymbol{\mu}_{1}^*,\boldsymbol{\mu}_{2}^*))$ satisfies the KKT conditions~\eqref{eq.kkt1}-\eqref{eq.kkt4}~\cite{BOYD}. 

Under the direct collocation approach, the cost and constraints are applied to the optimization variables $\boldsymbol{t_f}$, $\boldsymbol{x}(t)$, $I_0(t)$, and $\boldsymbol{I_B}(t)$ ($t_f$ is scalar in case of OCP-SCT) at collocation points (CPs). In this paper, we  determine the CPs based on the Gaussian quadrature formula (GQF) using which the BPs do not coincide with the CPs necessarily. 
GQF can find an optimal set of CPs (not equally spaced) to fit high-degree polynomials. After transcription of the OCP to the NLP problem using the direct collocation, the interior point solver IPOPT~\cite{IPOPT}
is employed to solve the NLP problem. All \textcolor{black}{the} dynamics, operating constraints, and the cost are 
\textcolor{black}{implemented} symbolically. This formulation
provides symbolic differentiation of the OCP, which in turn, results in remarkable improvement in convergence
time and solving 
\textcolor{black}{feasibility.} 

\begin{remark}
	In view of~\eqref{eq.NFP}, the number of free parameters reduces to 
	$N^{t}_{FP}=N_{FP_x}N_s+N_{FP_{I_B}}N_{cell}+N_{FP_{t_f}}+N_{FP_{I_0}}$	when OCP-SCT scheme is selected for the OCP. This results in the NLP with less parameters to be optimized with a reduction of computational effort and convergence time.     
\end{remark}

\section{Simulation Results}\label{section.Sim}
\begin{table}[t]
	\centering
	\caption{Specifications of the cylindrical 18650 LIB cell used in the simulations.}
	\label{table.spec}
	\centering
	\begin{tabular}{cc}
		\hline
		Model & Specification (Sony VTC4)\\
		\hline
		\hline
		\centering
		Cathode chemistry & NMC \\
		\centering
		Anode chemistry & Graphite \\
		\centering
		Nominal capacity  & 2 Ah \\
		\centering
		Nominal voltage & 3.6 V\\
		\centering
		Minimum voltage  & 2.5 V \\
		\centering
		Maximum voltage  & 4.2 V \\
	\end{tabular}
\end{table}
%
\begin{figure}[t]
	\scriptsize
	\begin{tabular}{cc}
		\hspace*{-1.3em}\includegraphics[scale=0.31]{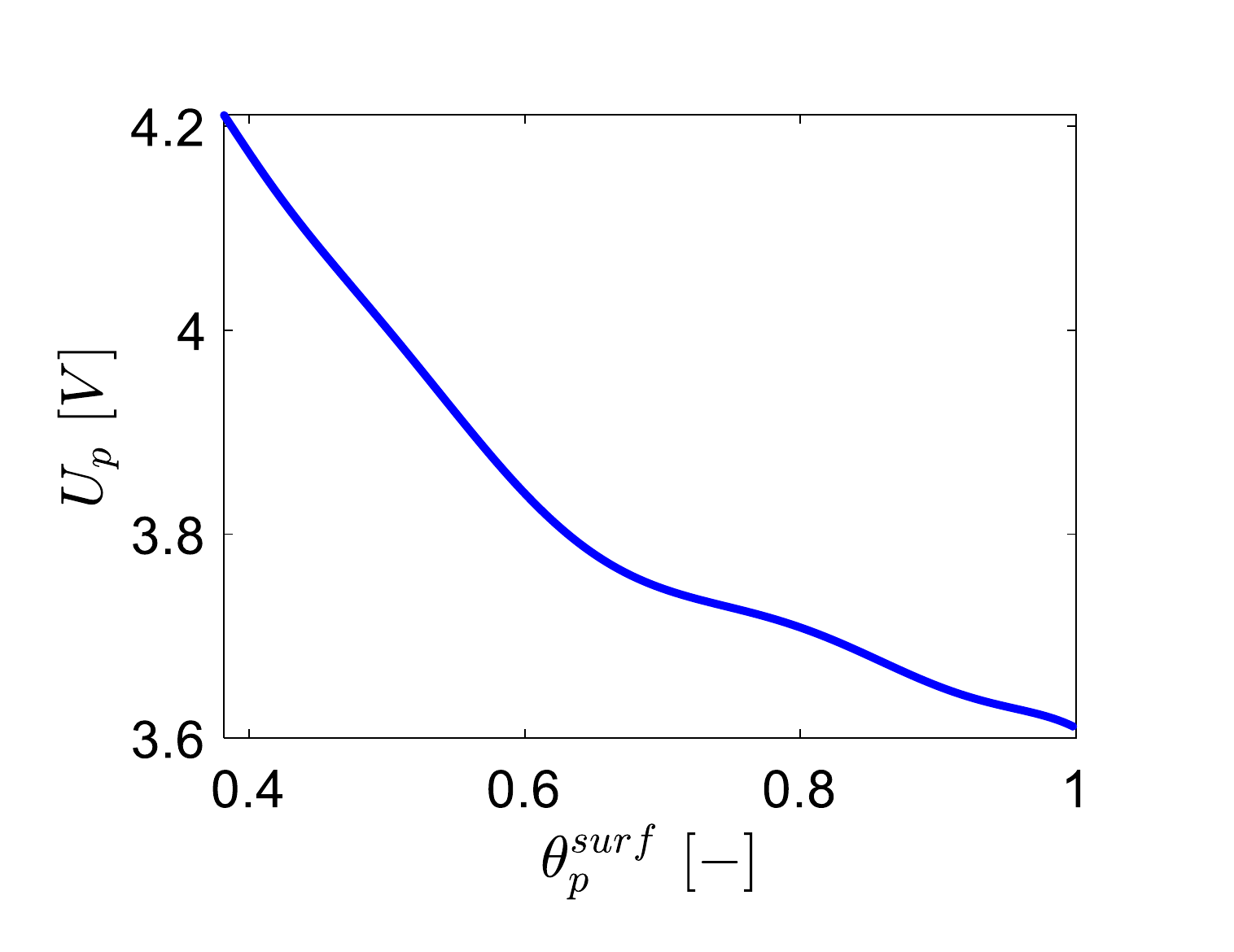}&
		\hspace*{-3.5em}\includegraphics[scale=0.31]{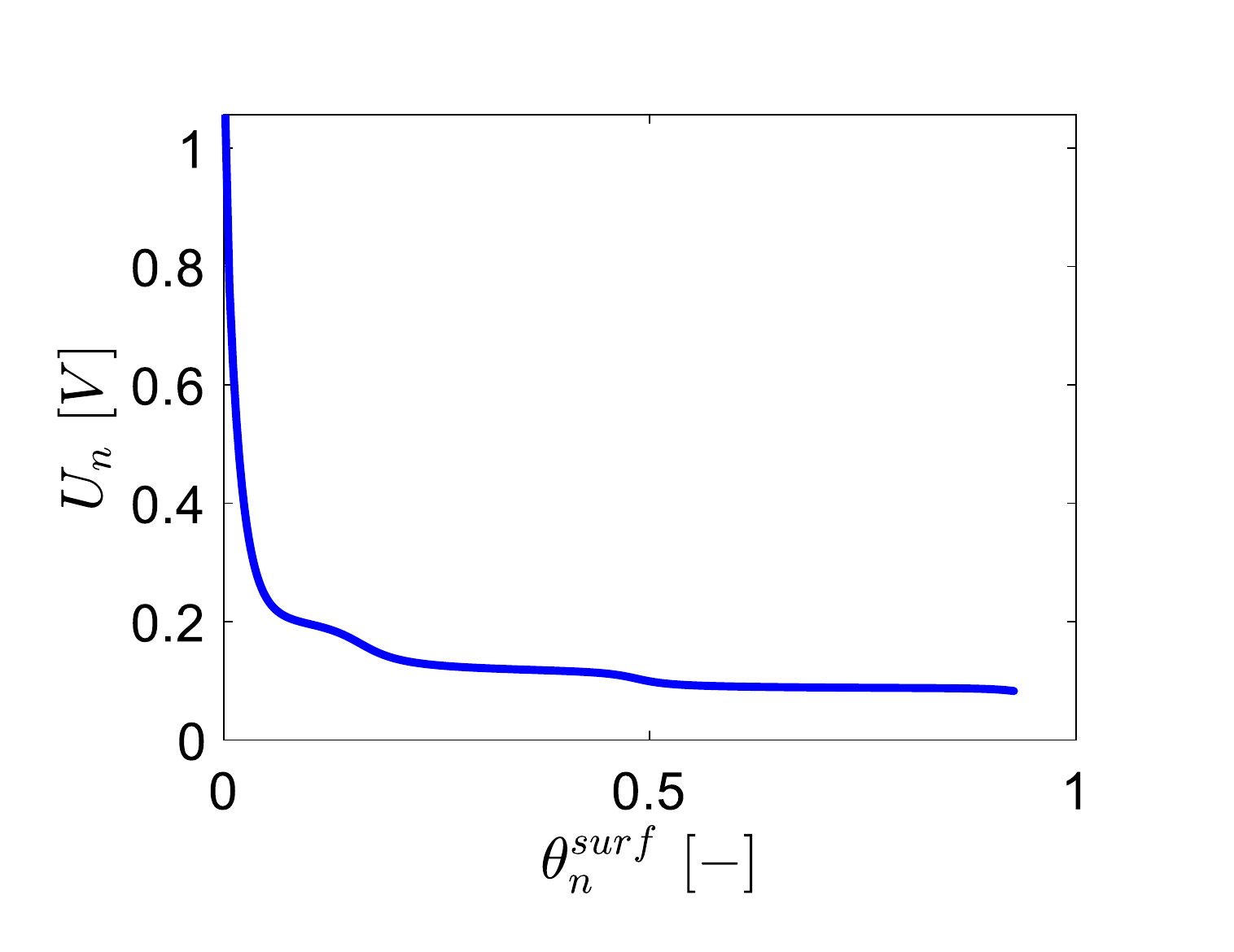}
	\end{tabular}
	\vspace{0.5em}
	\caption{Open-circuit potentials of NMC cathode/graphite anode cell.}
	\vspace{0em}
	\label{fig.OCP}
\end{figure}
%
\begin{figure*}[t]
	\scriptsize
	\begin{tabular}{ccc}
		\hspace*{-1.3em}\includegraphics[scale=0.43]{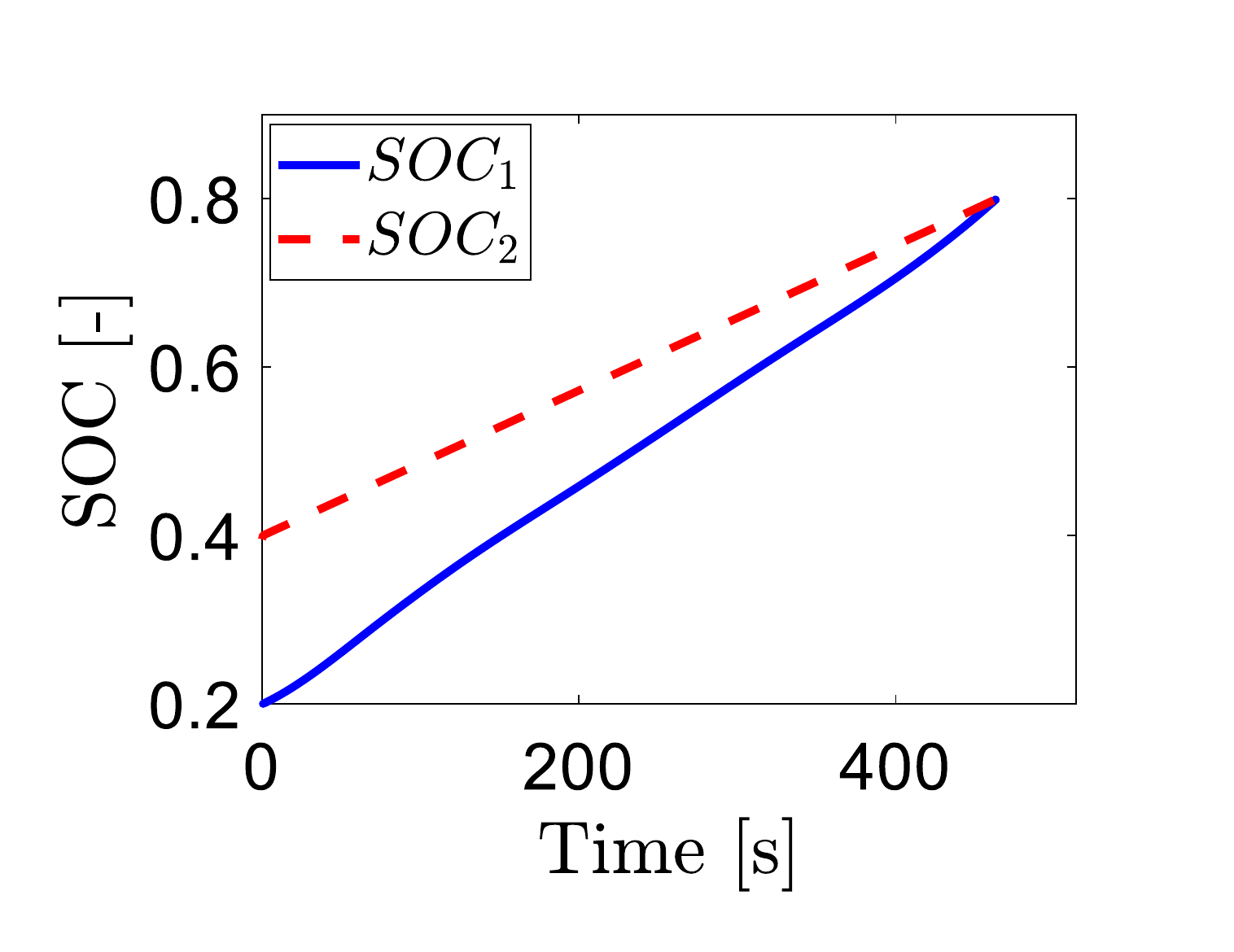}&
		\hspace*{-4em}\includegraphics[scale=0.43]{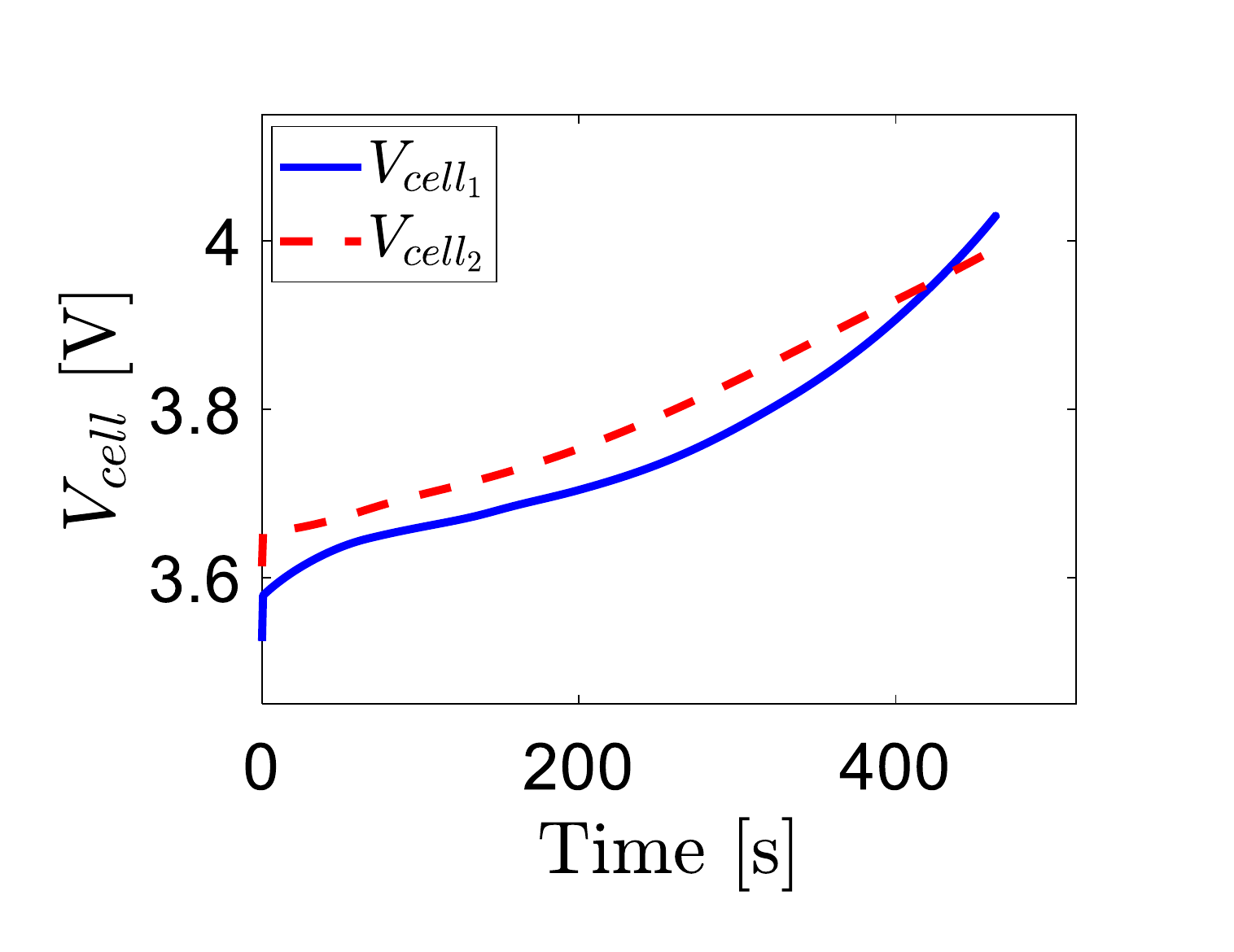} &
		\hspace*{-4em}\includegraphics[scale=0.43]{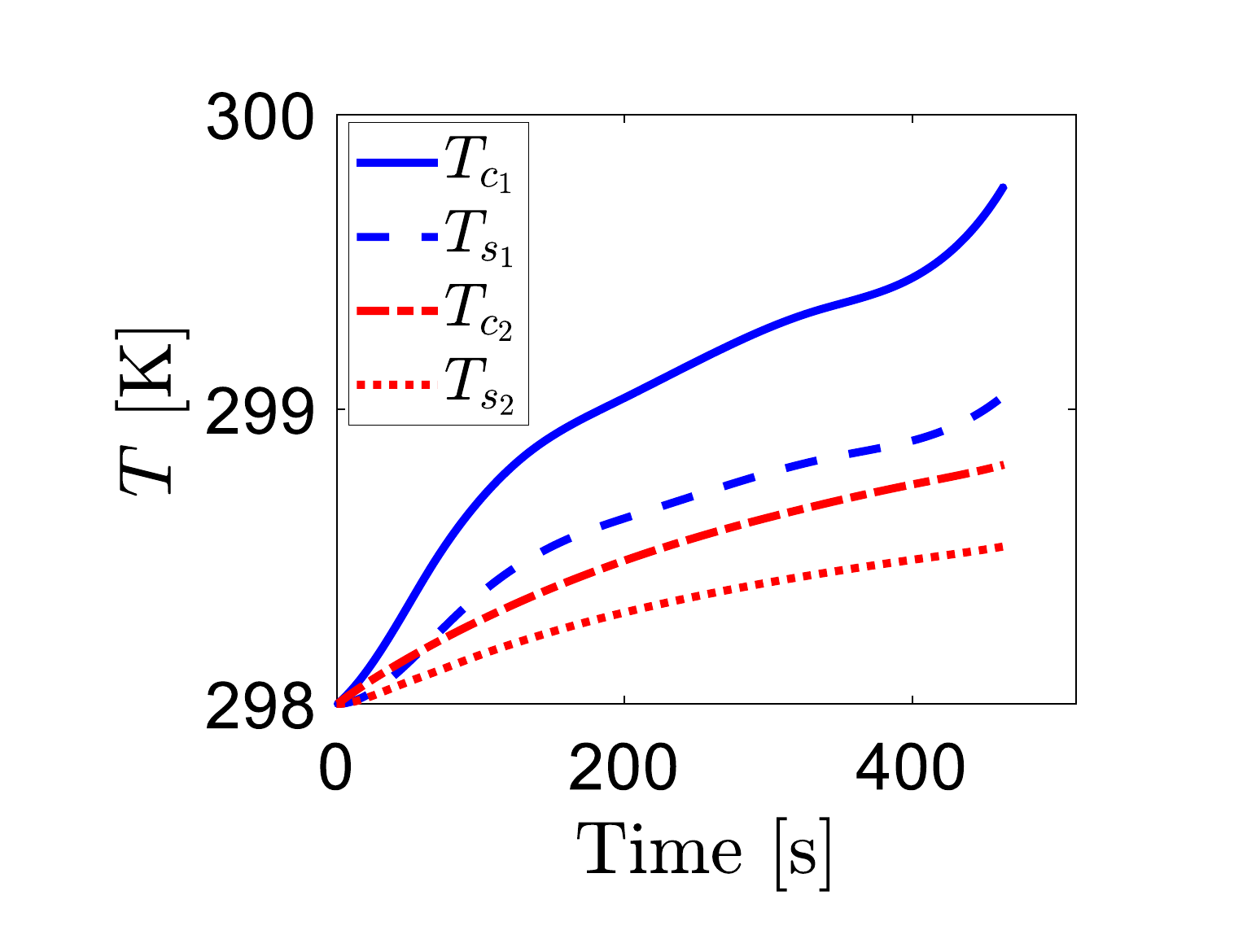} 
		\vspace*{-1.3em} \\
		\hspace*{-1.3em}\includegraphics[scale=0.43]{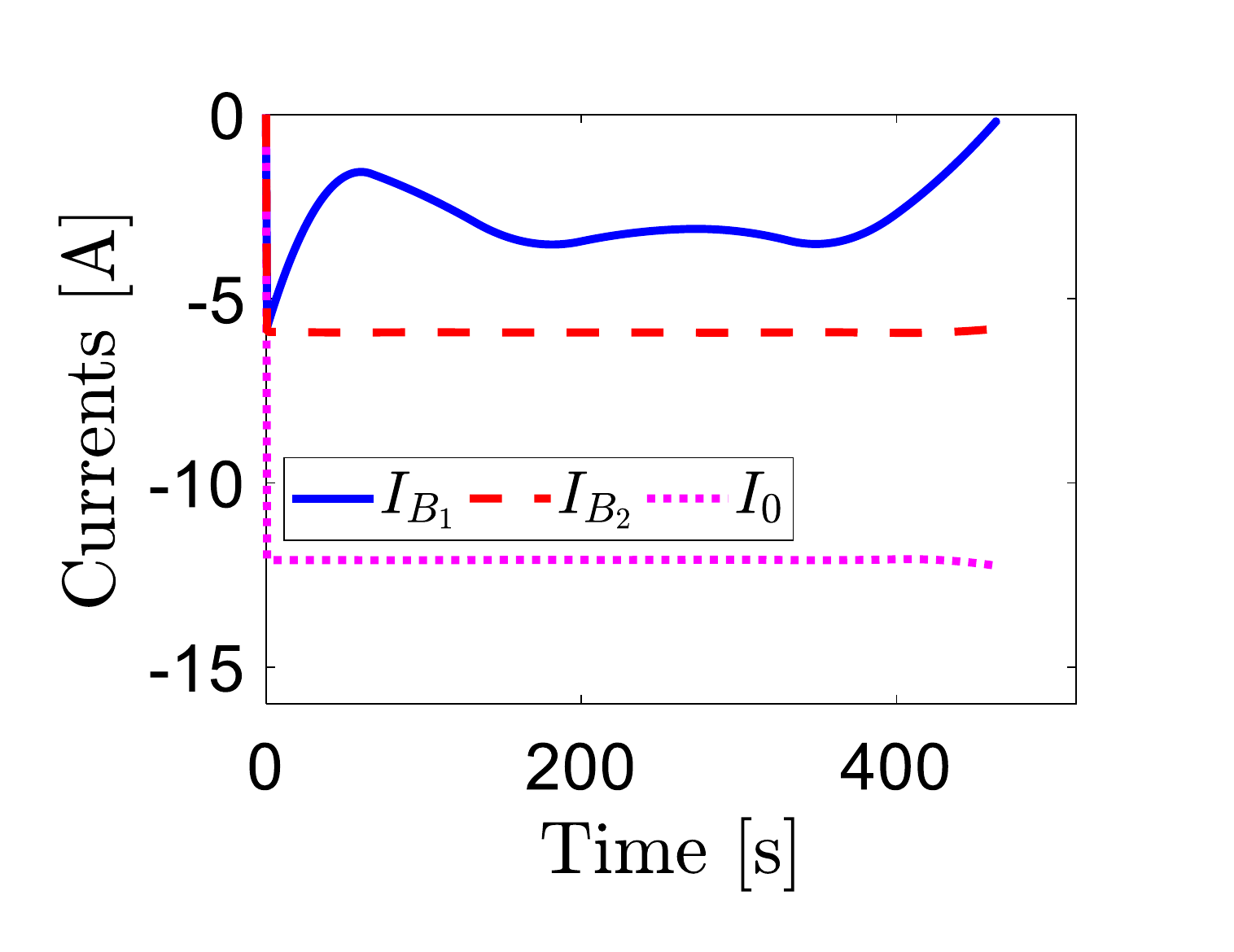}&
		\hspace*{-4em}\includegraphics[scale=0.43]{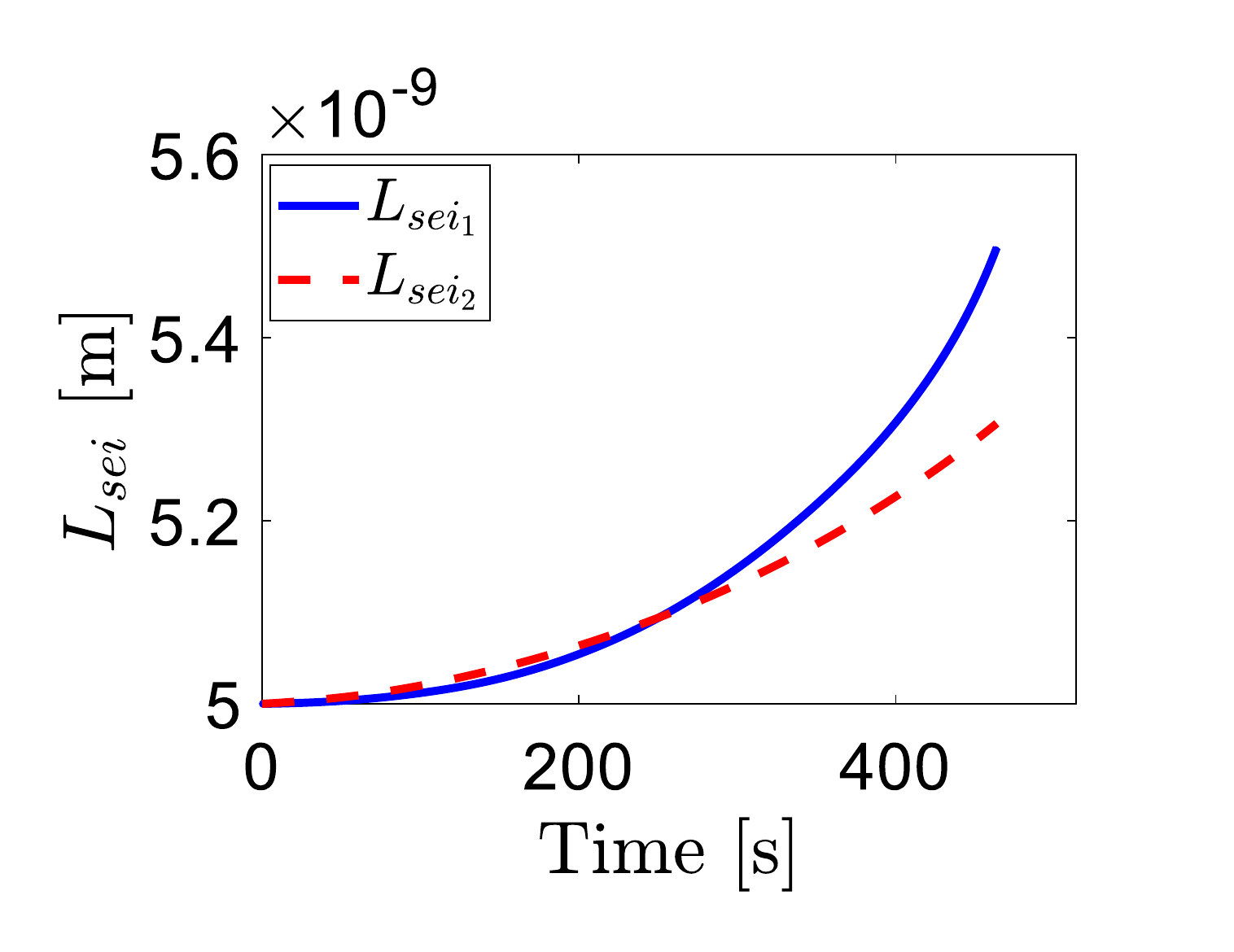}&
		\hspace*{-4em}\includegraphics[scale=0.43]{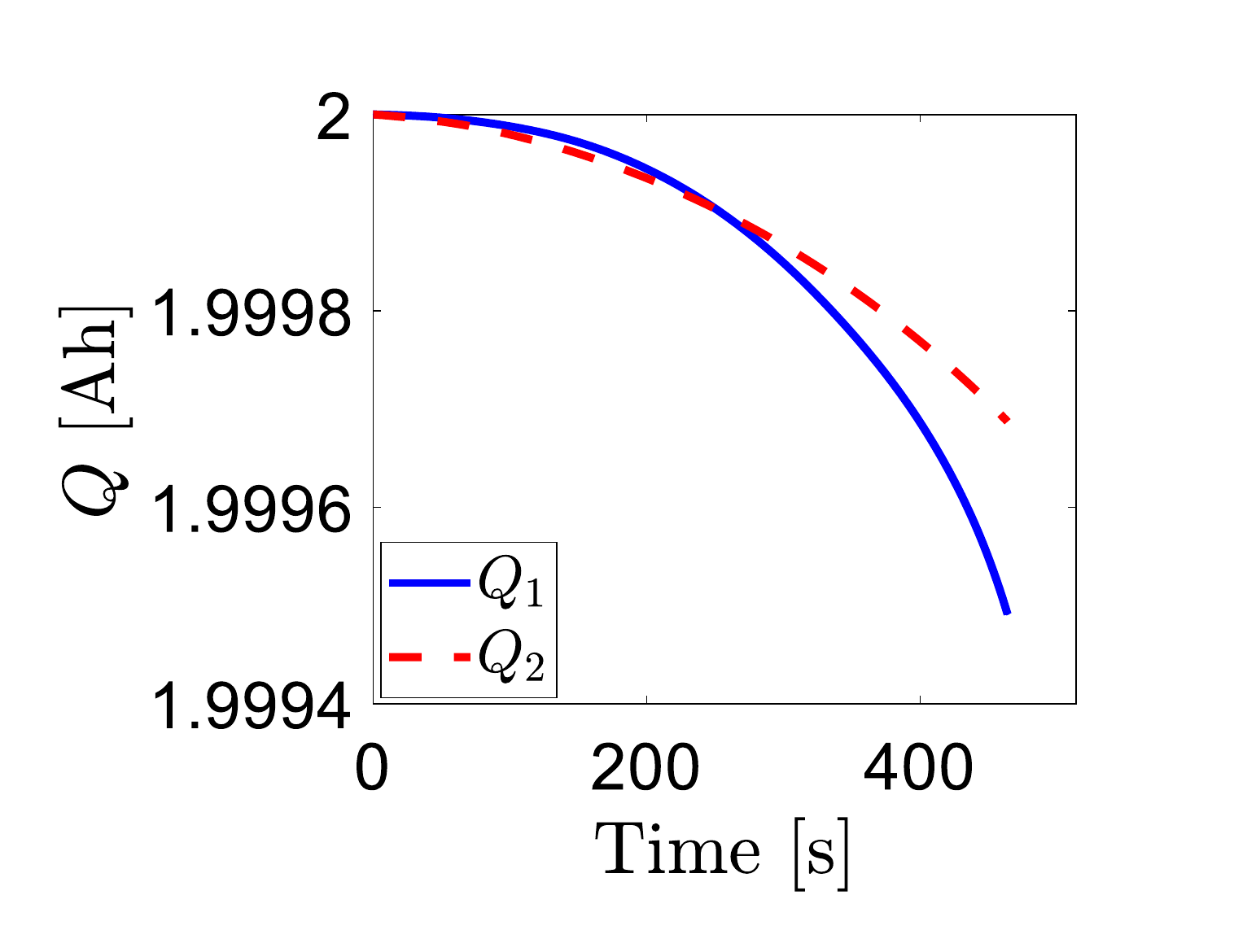}
	\end{tabular}
	\vspace{0em}
	\caption{Results from the multi-objective OCP-SCT scheme with initial SOC mismatch $SOC(0)=[0.2, 0.4]$ at $T_{amb}=25^{\circ} \text{C}$.}
	\vspace{-0em}
	\label{fig.SCT25}
\end{figure*}

\begin{figure*}[t]
	\scriptsize
	\begin{tabular}{ccc}
		\hspace*{-1.3em}\includegraphics[scale=0.43]{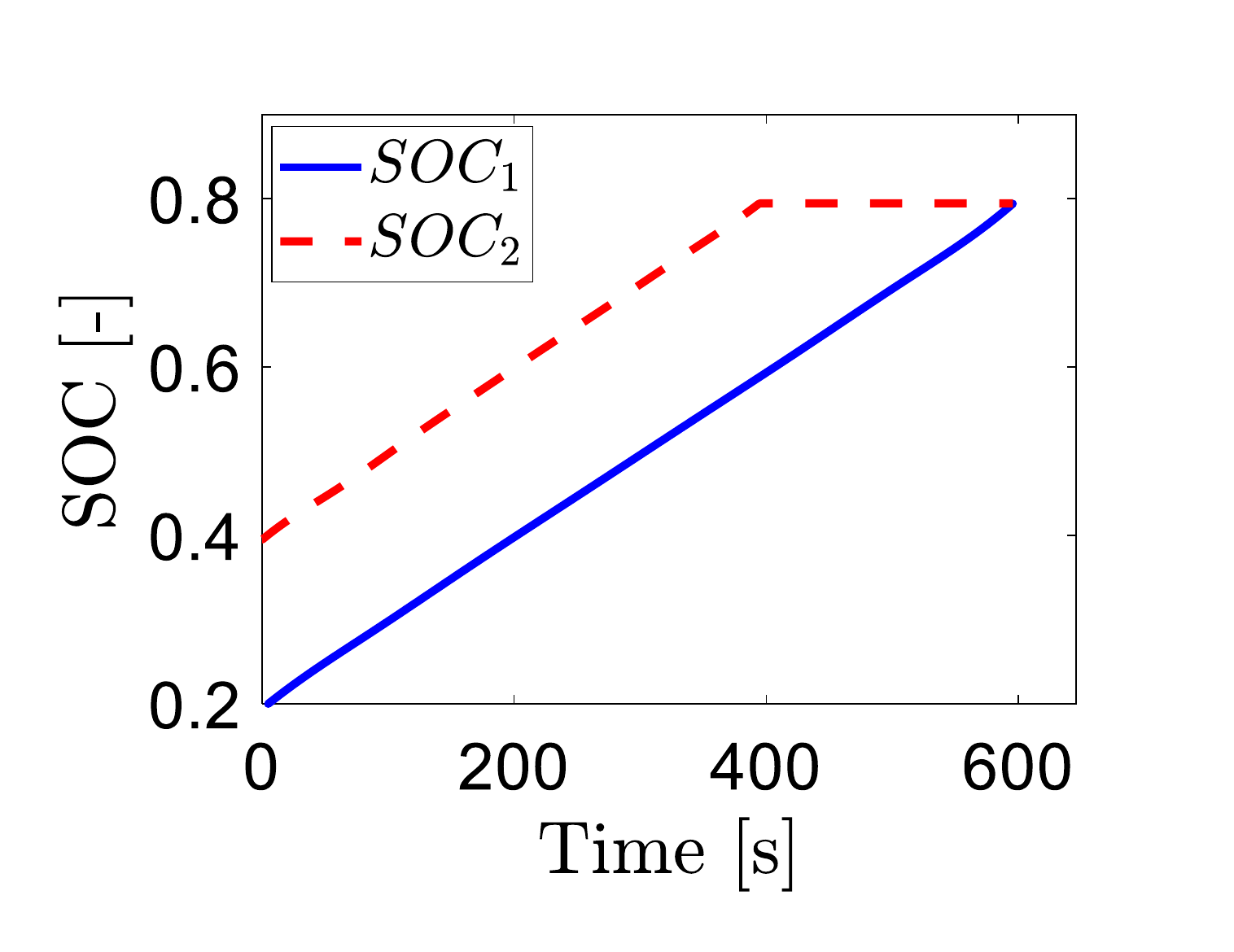}&
		\hspace*{-4em}\includegraphics[scale=0.43]{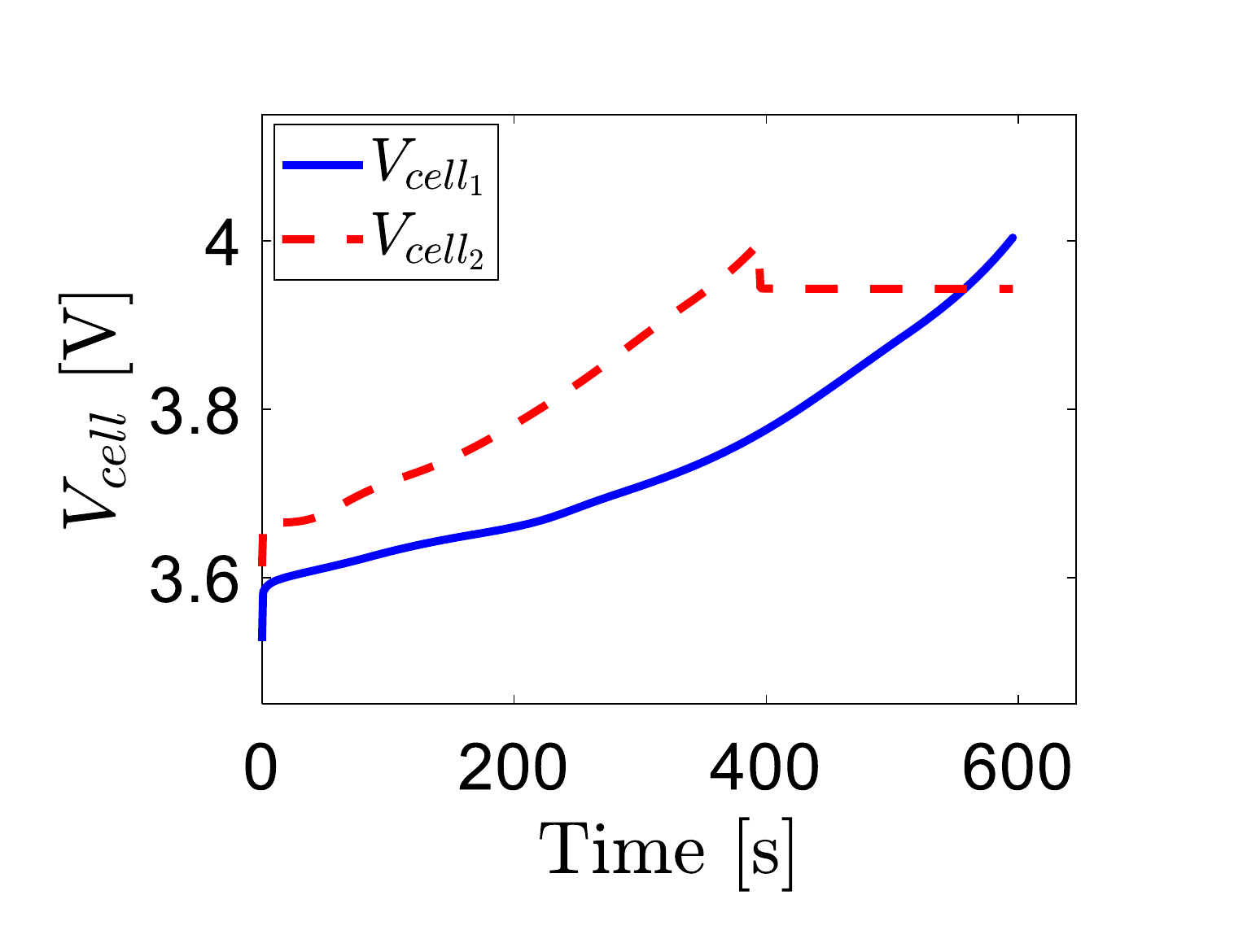} &
		\hspace*{-4em}\includegraphics[scale=0.43]{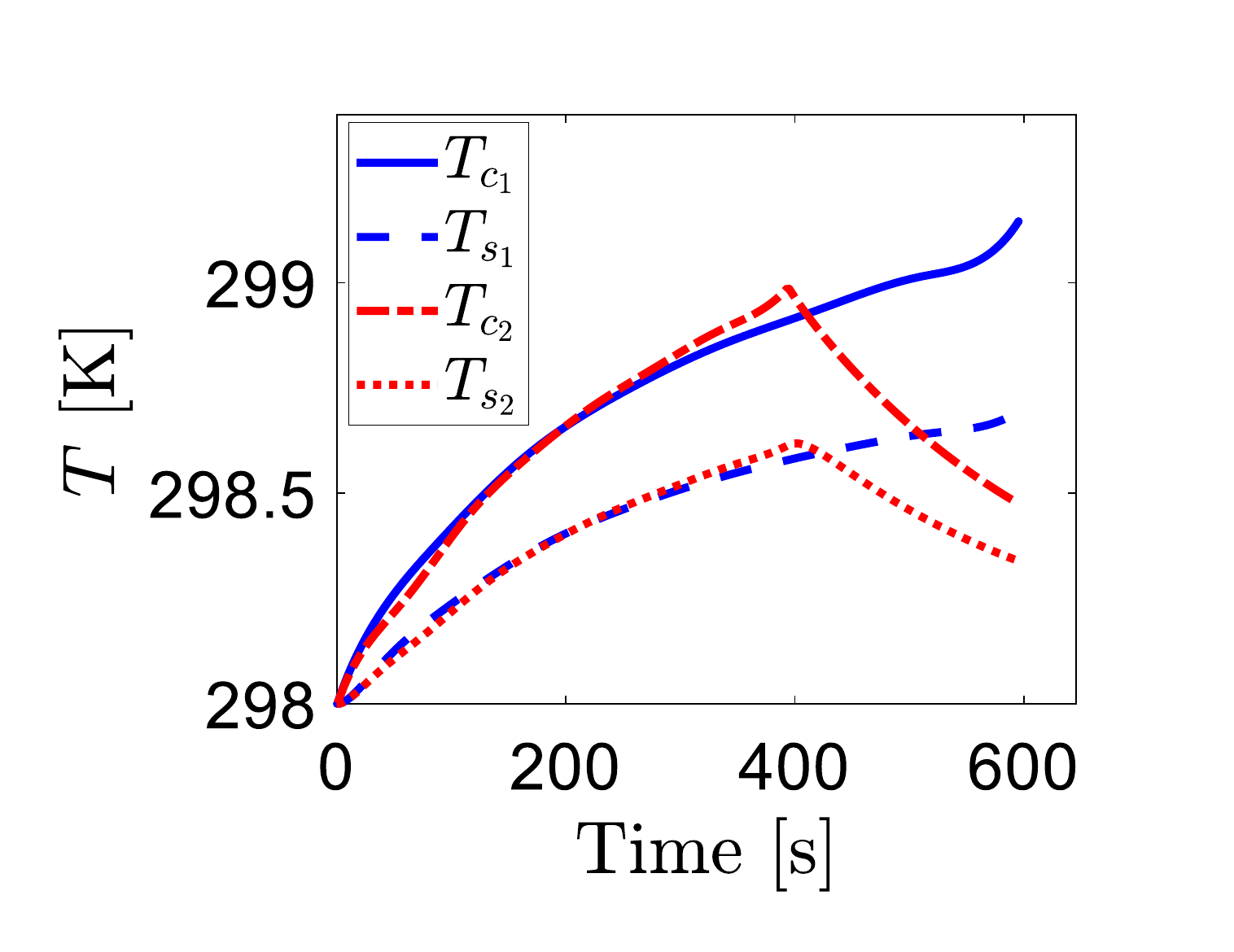} 
		\vspace*{-1.3em} \\
		\hspace*{-1.3em}\includegraphics[scale=0.43]{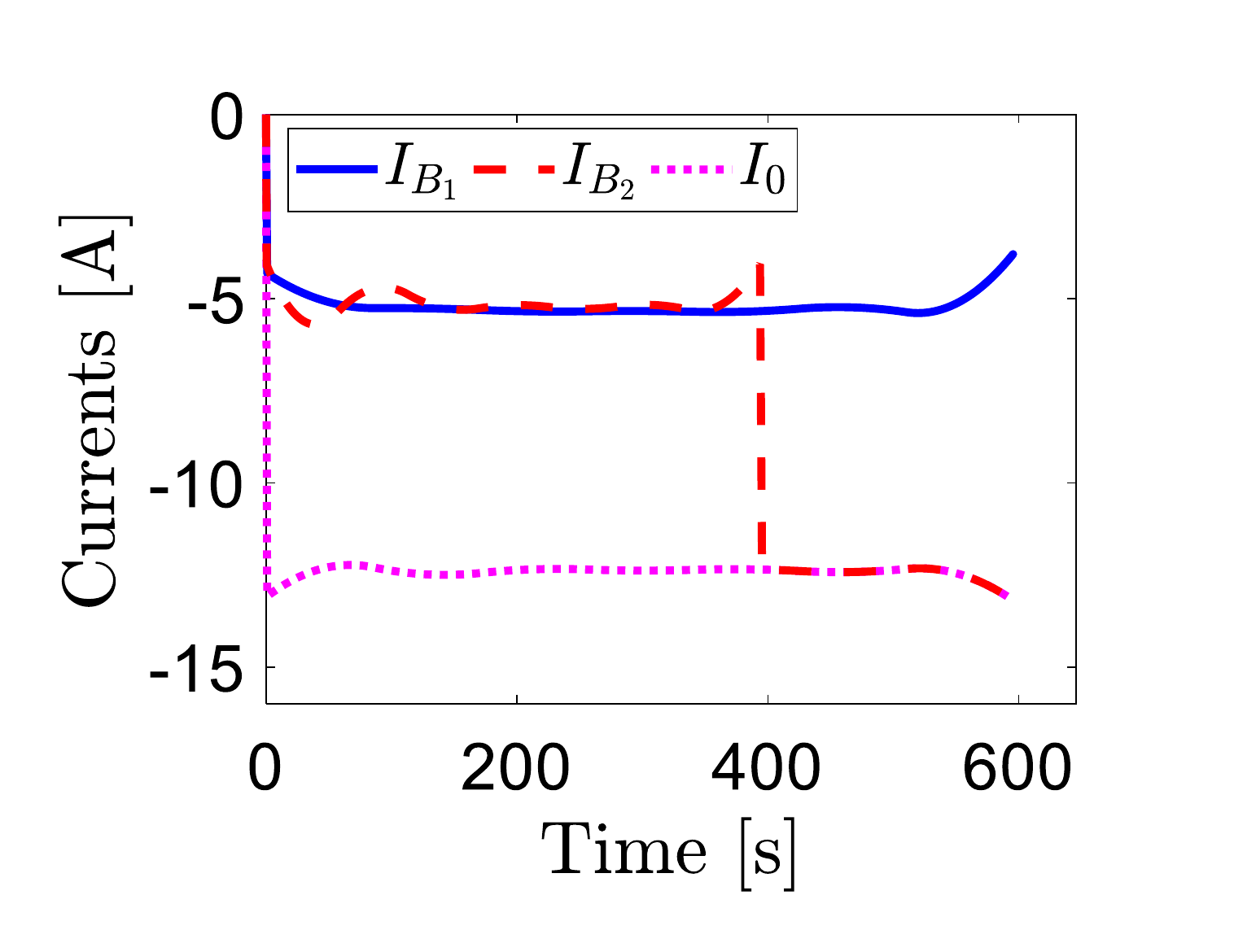}&
		\hspace*{-4em}\includegraphics[scale=0.43]{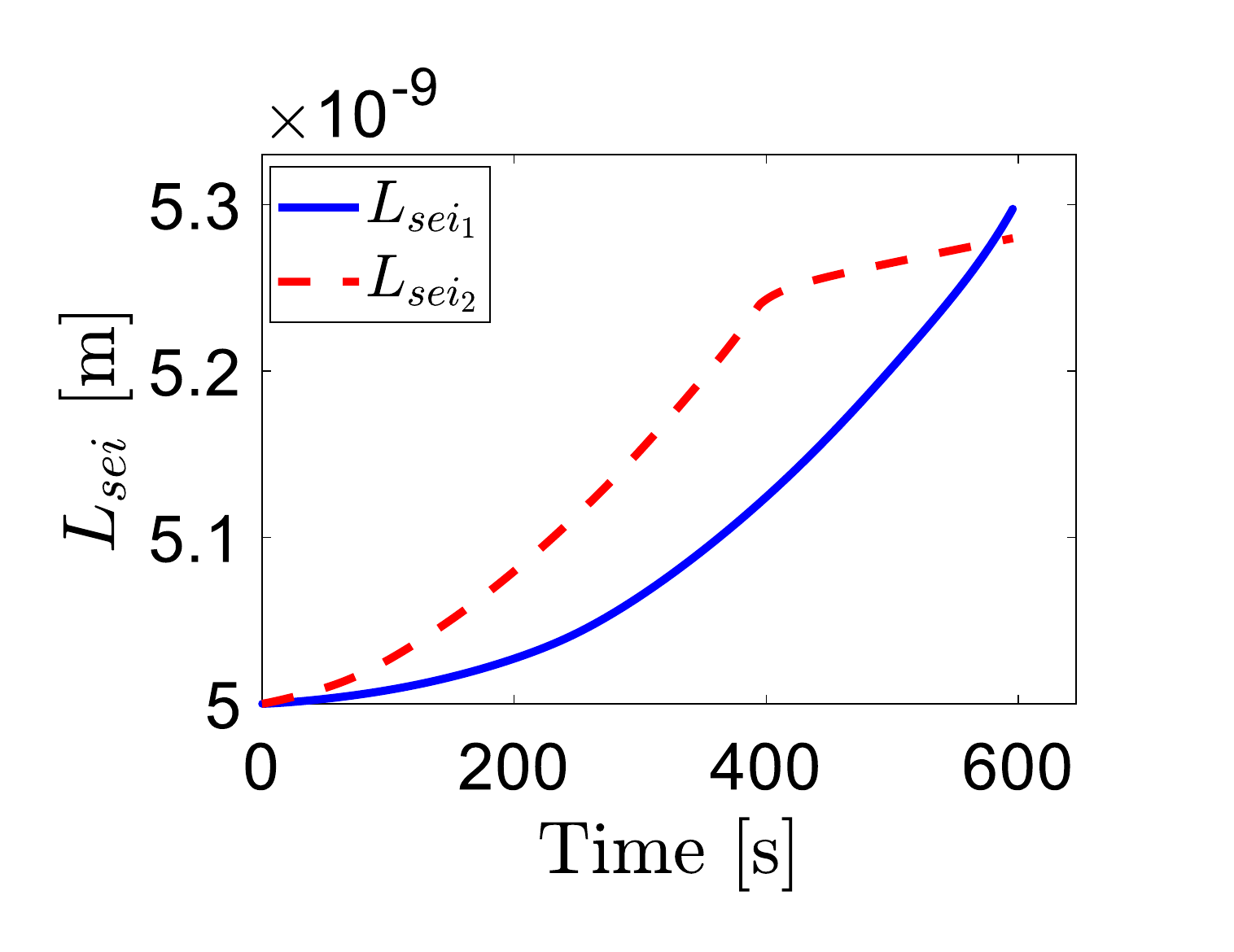}&
		\hspace*{-4em}\includegraphics[scale=0.43]{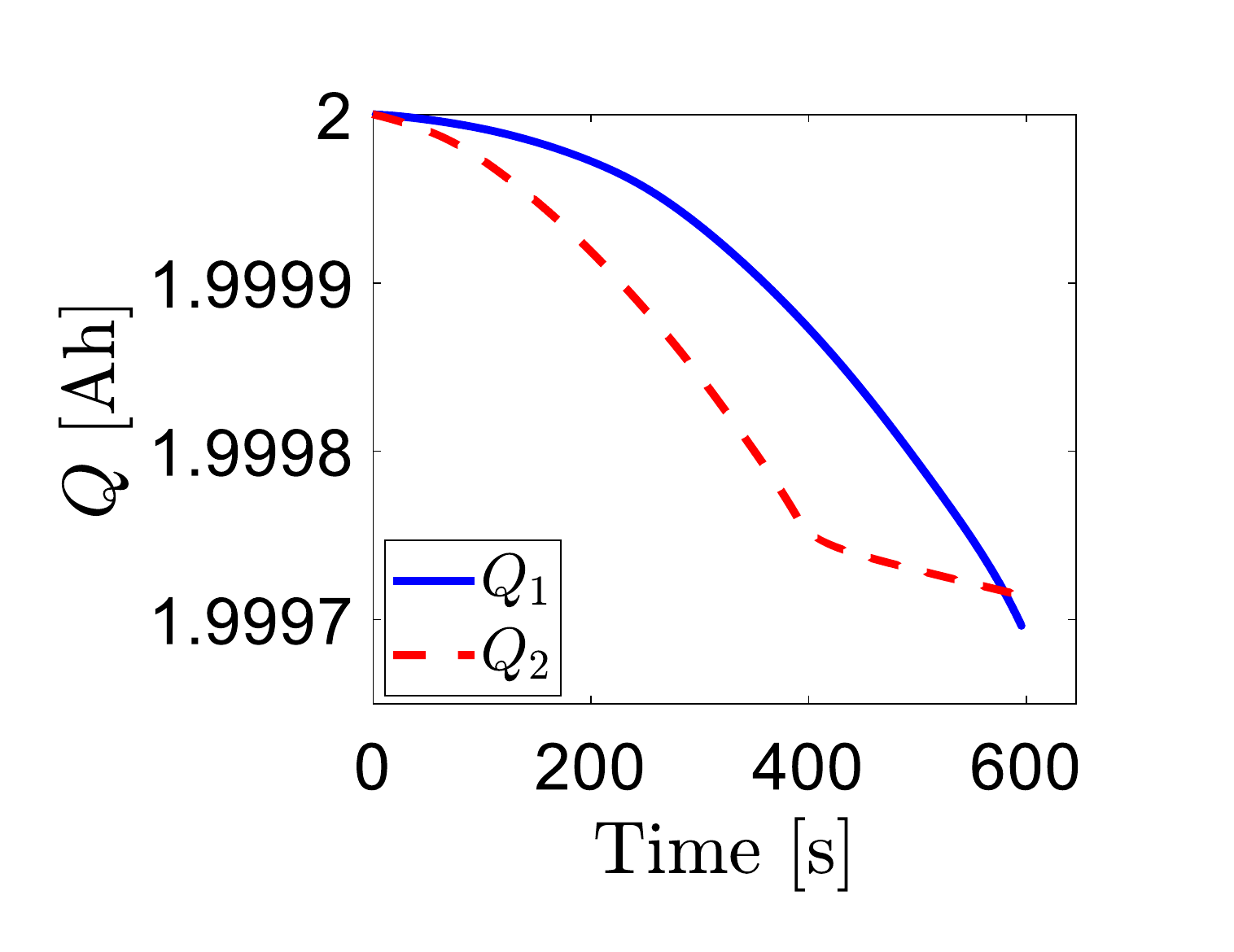}
	\end{tabular}
	\vspace{0em}
	\caption{Results from the multi-objective OCP-DCT scheme with initial SOC mismatch $SOC(0)=[0.2, 0.4]$ at $T_{amb}=25^{\circ} \text{C}$.}
	\vspace{-0em}
	\label{fig.DCT25}
\end{figure*}



In this section, we test the effectiveness of the proposed optimal control algorithm for both OCP-SCT and OCP-DCT schemes on a battery module with two series connected imbalanced cells (i.e., $N_{cell}=2$), where each cell is connected in parallel to an active balancing circuitry  (see Fig.~\ref{fig.Pack5}). 

\begin{table*}[t]
	\centering
	\caption{Performance comparison between the OCP-SCT and OCP-DCT schemes with initial SOC mismatch $SOC_1(0)=0.2$ and  $SOC_2(0)=0.4$ at different ambient temperatures $T_{amb}=[15, 25, 35]^{\circ} C$.
		The best value of each metric is shown in bold.}
	\label{table.Ex1}
	\centering
	\begin{tabular}{ccccc}
		\hline
		$T_{amb}$ & Scheme & $(\Delta L_{sei_1}, \Delta L_{sei_2})\%$ 
		& $(\Delta Q_{1}, \Delta Q_{2})\%$ & $(t_{f_1}, t_{f_2})\ \text{s}$\\
		\hline
		\hline
		\centering
		\multirow{2}{*}{$15^{\circ} C$} &OCP-SCT & (0.079, \textbf{0.021}) & (20, 5.5)$\times$10$^{-5}$ & (\textbf{406}, 406)\\
		\centering
		&OCP-DCT & (\textbf{0.021}, \textbf{0.021}) & (\textbf{5.43}, \textbf{5.41})$\times$10$^{-5}$ & (571, \textbf{386}) \\
		\hline 
		\centering
		\multirow{2}{*}{$25^{\circ} C$} &OCP-SCT & (9.97, 6.13) & (0.025, 0.015) & (\textbf{463}, 463)\\
		\centering
		&OCP-DCT & (\textbf{5.95}, \textbf{5.59}) & (\textbf{0.015}, \textbf{0.014}) & (595, \textbf{393}) \\
		\hline 	
		\centering
		\multirow{2}{*}{$35^{\circ} C$} &OCP-SCT & (60, 47) & (0.15, 0.12) & (\textbf{471}, 471)\\
		\centering
		&OCP-DCT & (\textbf{38}, \textbf{29}) & (\textbf{0.098}, \textbf{0.073}) & (671, \textbf{458}) 	
	\end{tabular}
	\vspace{-1em}
\end{table*}

\begin{figure*}
	\begin{tabular}{cc}
		\hspace*{-1.5em}\includegraphics[scale=0.27]{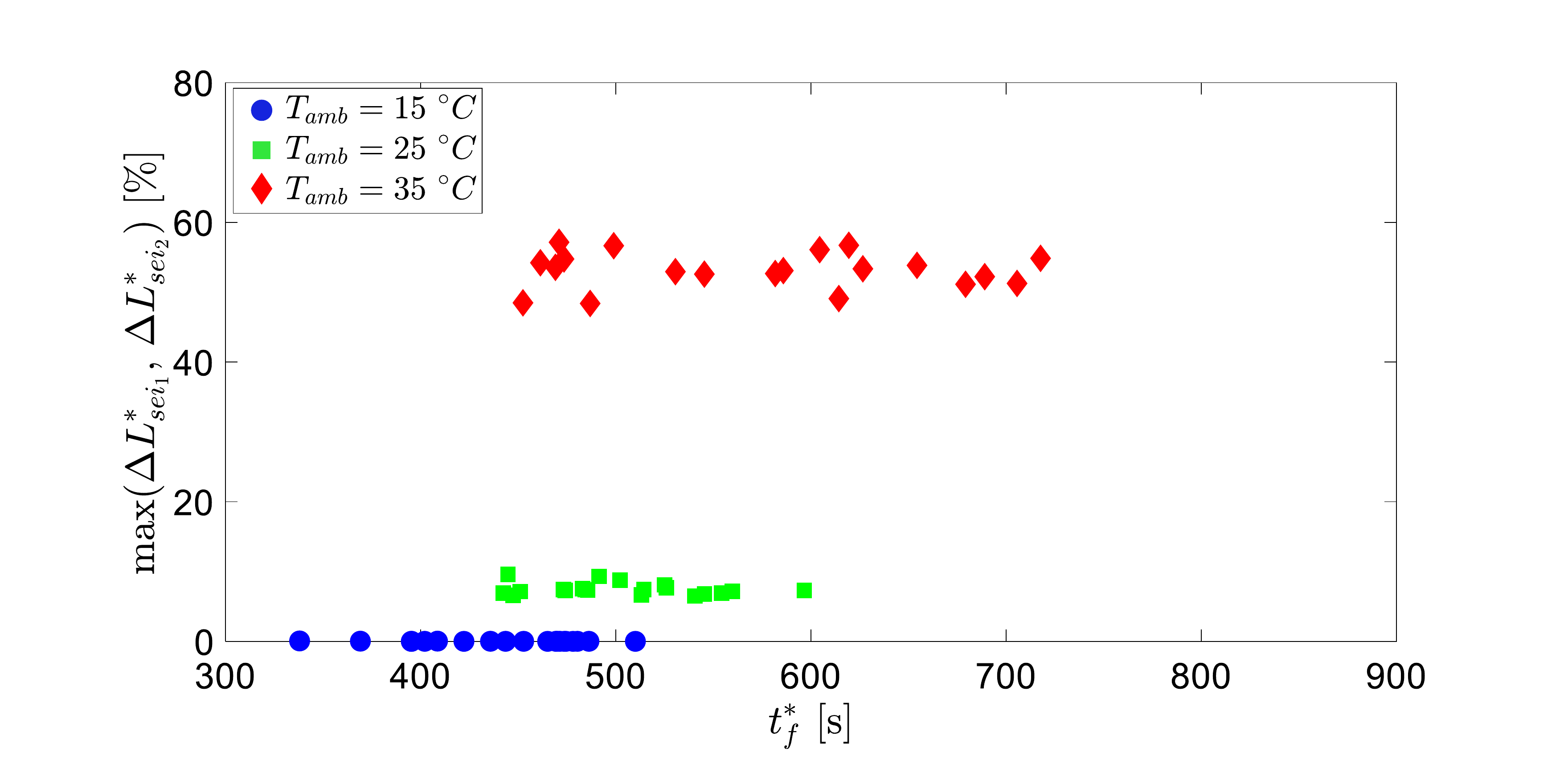}&
		\hspace*{-3.5em}\includegraphics[scale=0.27]{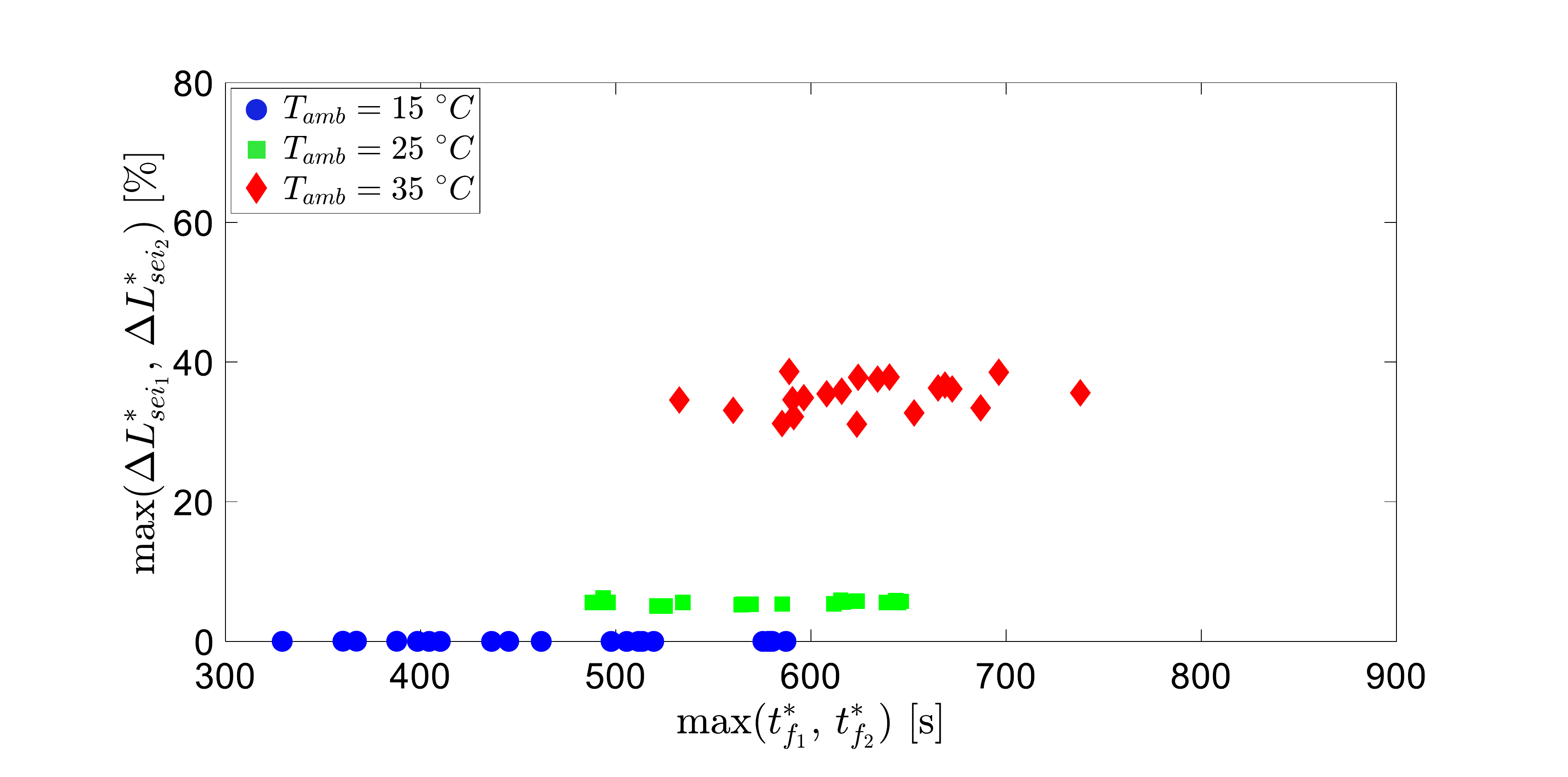}\\
		\hspace*{-4.5em}  {\small (a) OCP-SCT scheme} & \hspace*{-5.5em} {\small (b) OCP-DCT scheme}
	\end{tabular}
	\vspace{1em}
	\caption{Optimization results for the OCP-SCT and OCP-DCT schemes based on 20  simulations with initial SOC mismatch taken from a uniform distribution over the interval $[0.2, 0.4]$  at different ambient temperatures $T_{amb}=[15, 25, 35]^{\circ} \text{C}$.} 
	\vspace{-1em}
	\label{fig.MC-SOC}
\end{figure*}

\begin{figure*}
	\begin{tabular}{cc}
		\hspace*{-1.5em}\includegraphics[scale=0.27]{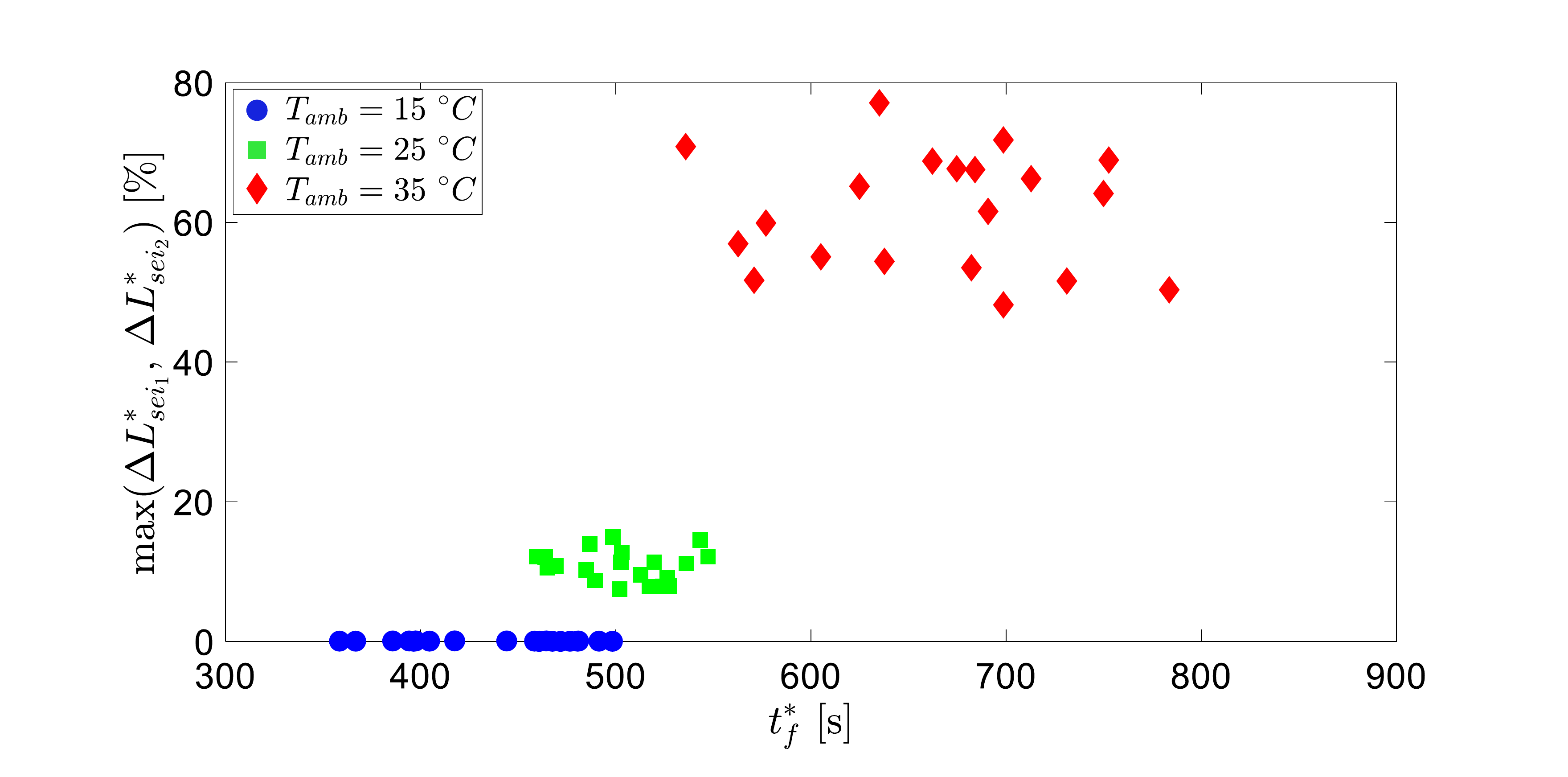}&
		\hspace*{-3.5em}\includegraphics[scale=0.27]{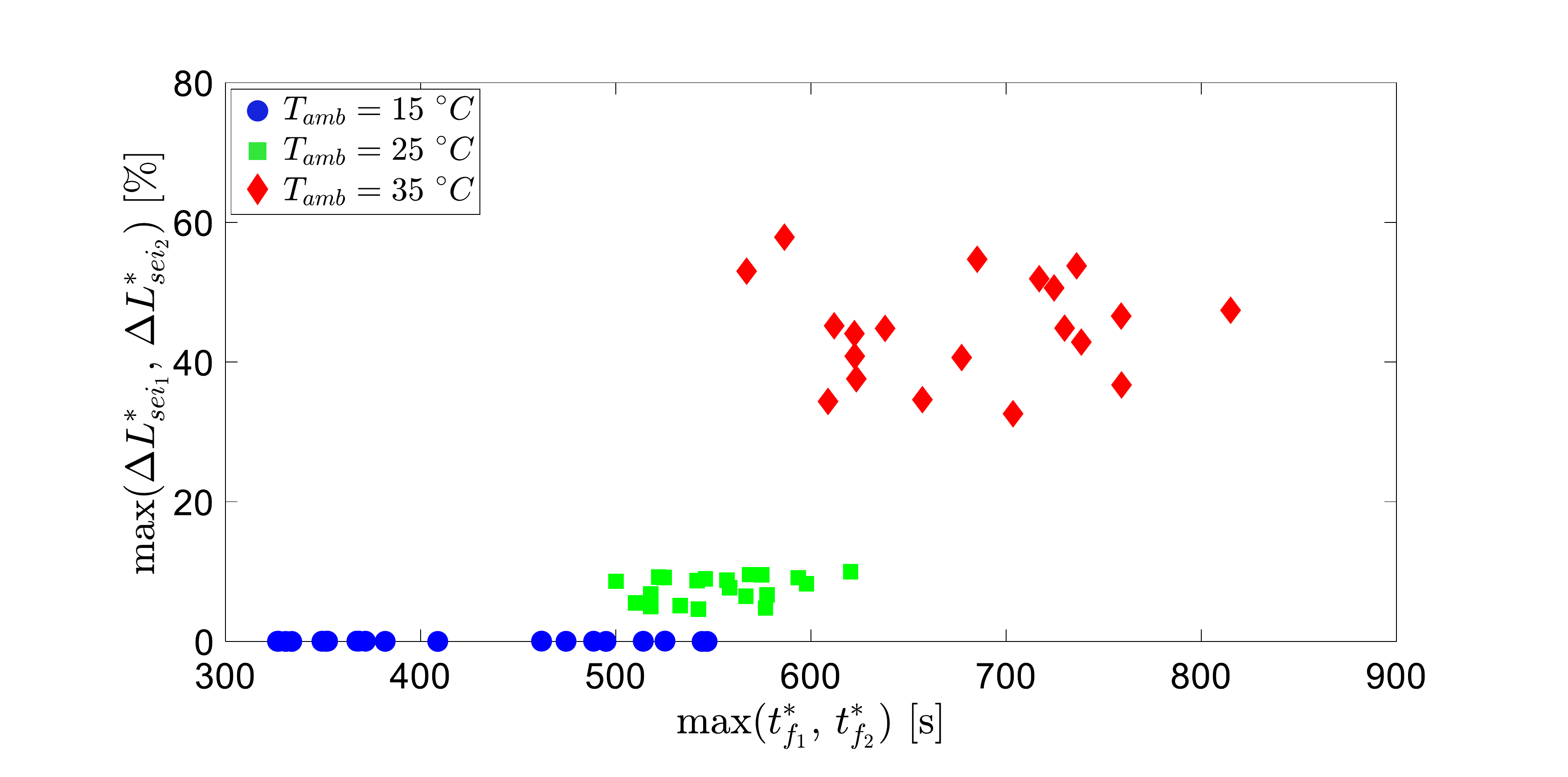}\\
		\hspace*{-4.5em}  {\small (a) OCP-SCT scheme} & \hspace*{-5.5em} {\small (b) OCP-DCT scheme}
	\end{tabular}
	\vspace{1em}
	\caption{Optimization results for the OCP-SCT and OCP-DCT schemes based on 20 simulations with initial $L_{sei}$ mismatch taken from a uniform distribution over the interval $[4, 6]\times10^{-9} m$  at different ambient temperatures $T_{amb}=[15, 25, 35]^{\circ} \text{C}$.} 
	\vspace{-1em}
	\label{fig.MC-Lsei}
\end{figure*}

\subsection{Initialization and set up}

The battery considered in this paper is a cylindrical 18650, 2-Ah lithium-ion nickel–manganese–cobalt (NMC) cathode/graphite anode cell
whose characteristics are reported in Table~\ref{table.spec}~\cite{Allam2019}. The open-circuit potentials of each electrode, $U_j$, in terms of the surface stoichiometry, $\theta^{surf}_j=c^{surf}_{s,j}/c^{max}_{s,j}$, is illustrated in Fig.~\ref{fig.OCP}. Throughout the simulations, we assume that there is an initial SOC imbalance  among the cells ($SOC_1(0)\neq SOC_2(0)$) while no mismatch between temperature, SEI layer thickness, resistance, and capacities of individual battery cells is assumed.

The physical bounds for the operating constraints~\eqref{eq.const-I}-\eqref{eq.const-t} are set to $I_{B_{\textrm{min}}}=-6 \ \text{A}$, $I_{0_{\textrm{min}}}=-16 \ \text{A}$, $I_{B_{\textrm{max}}}=0 \ \text{A}$, $I_{0_{\textrm{max}}}=-12 \ \text{A}$, $SOC_{\textrm{target}}=0.8$, $V_{cell{\textrm{min}}}=2.5 \ \text{V}$, $V_{cell{\textrm{max}}}=4.2 \ \text{V}$, $t_{f_{\textrm{max}}}=2000 \ \text{s}$, $T_{lk_{\textrm{min}}}=5^{\circ}\text{C}$, and $T_{lk_{\textrm{max}}}=45^{\circ} \text{C}$ with $l\in\{c,s\}$ and $k=1,2$. Note that the minimum and maximum voltages follow the battery specifications mentioned in Table~\ref{table.spec}. The initial conditions are picked as $L_{sei_k}(0)=5\times10^{-9} \ \text{m}$ (this is the typical SEI later thickness observed for a fresh cell), $Q_{k}(0)=2 \ \text{Ah}$, and $T_{ck}(0)=T_{sk}(0)=T_{amb}$.
The numbers of discretization points are set to $N_{r,j}=N_{sei}=10$. Thus, the number of states used in the OCP is $N_s=44$.
Given the nominal capacity $Q_{nom}=Q_{k}(0)=2 \ \text{Ah}$, the selected bounds for module and balancing currents result 
\textcolor{black}{in having cell current} 
\textcolor{black}{between $-16$A and $-6$A.} For the \textit{balanced charging-degradation scenario}, the optimization weights and the \textcolor{black}{trade-off coefficients} are selected to be $\beta_1=1$ $[s^{-1}]$, $\beta_2=\beta_3=5\times10^{8}$ $[\text{sm}^{-1}]$, and $\alpha=0.5$; they are the same for both the OCP-SCT and the OCP-DCT schemes.   

For the surrogate model development, the cell current is discretely sampled within its range, i.e., $[-16 \ -6]\ \text{A}$, (with sampling current 2 $A$) and ambient temperatures are chosen to be 
\textcolor{black}{$[15, 25, 35]^{\circ} \text{C}$. $5^{th}$-order polynomials} are fitted to the optimal points \textcolor{black}{$c_{solv}^{surf*}$} for all six sampled currents and each ambient temperature. The MATLAB built-in functions \textit{fminsearch} and \textit{polyfit} are employed to solve the optimization~\eqref{eq.Csolve} and fit the polynomials, respectively. 

\begin{figure*}
	\begin{tabular}{cc}
		\hspace*{-1.5em}\includegraphics[scale=0.27]{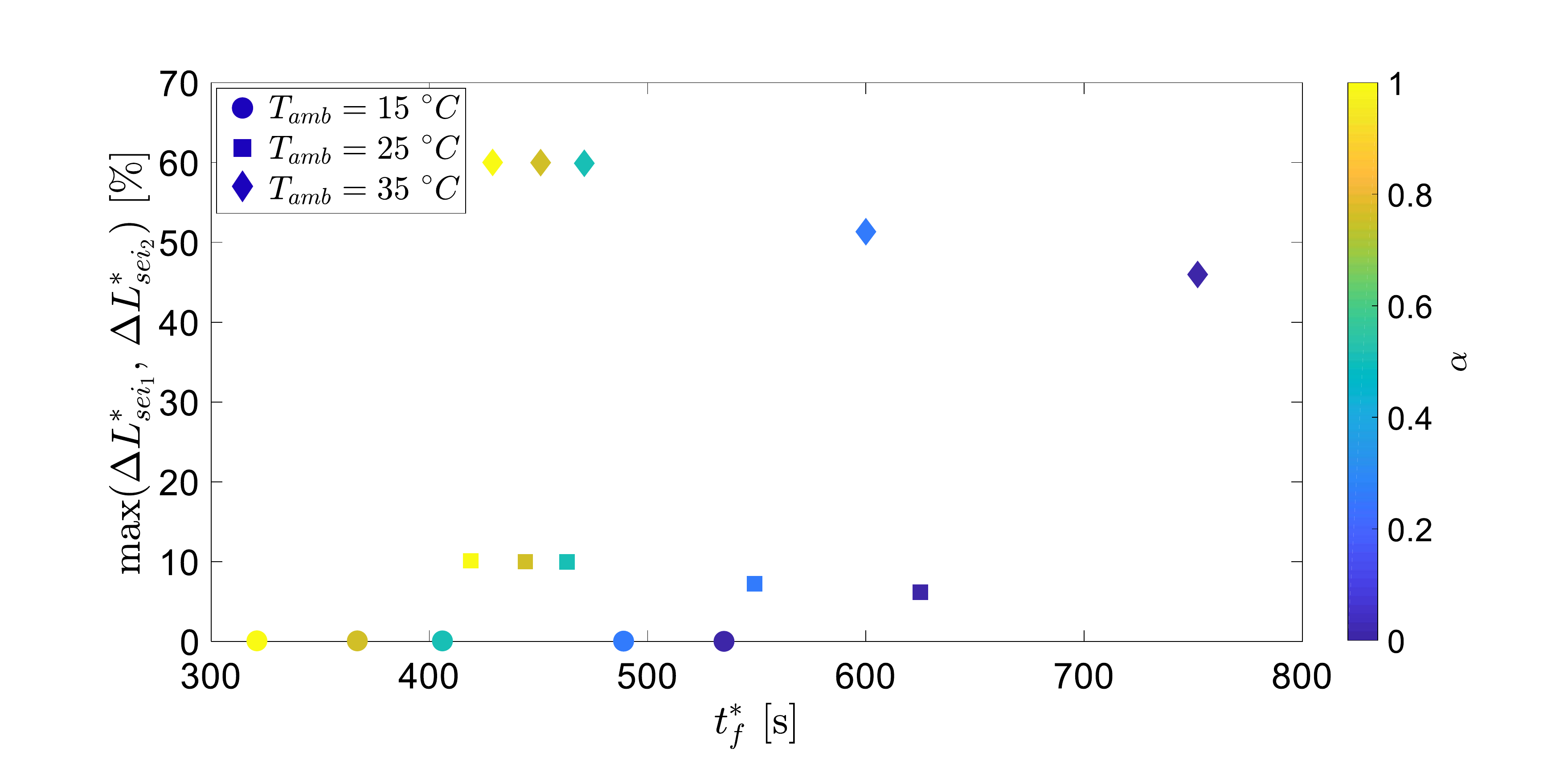}&
		\hspace*{-3.5em}\includegraphics[scale=0.27]{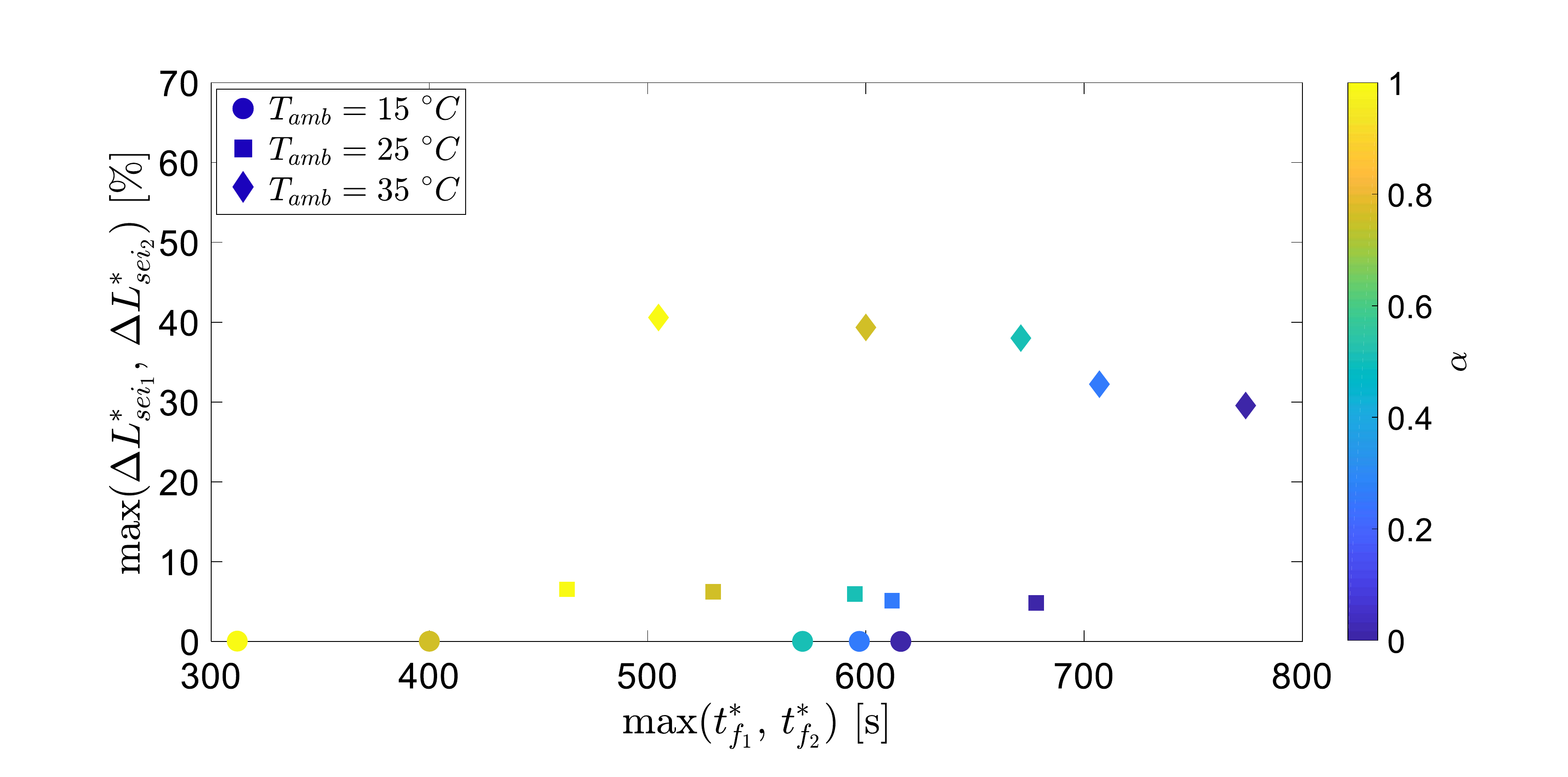}\\
		\hspace*{-4.5em}  {\small (a) OCP-SCT scheme} & \hspace*{-5.5em} {\small (b) OCP-DCT scheme}
	\end{tabular}
	\vspace{1em}
	\caption{Pareto fronts for the OCP-SCT and OCP-DCT schemes at different ambient temperatures $T_{amb}=[15, 25, 35]^{\circ} \text{C}$ when the optimization trade-off coefficient is discretely sampled as $\alpha=\{0,\ 0.25,\ 0.5,\ 0.75, 1\}$ and with an initial SOC mismatch $SOC(0)=[0.2, 0.4]$.} 
	\vspace{-1em}
	\label{fig.Pareto}
\end{figure*}

\subsection{Initial SOC mismatch with different ambient temperatures}
At an ambient temperature of $T_{amb}=25^{\circ} \text{C}$, we first solve OCP-SCT and OCP-DCT 
\textcolor{black}{when the initial SOC for the two cells is set to $SOC(0)=[0.2, 0.4]$}.
From Fig.~\ref{fig.SCT25}, under OCP-SCT scheme,  both cells are charged simultaneously while their voltages lie within $V_{cell{\textrm{min}}}=2.5 $V and $V_{cell{\textrm{max}}}=4.2 $V.  
In Cell 1 (with lower initial SOC) there is a higher rate of charge than in Cell 2 and the same time of charge is enforced. This, in turn, leads to Cell 1 to experience more aging, and achieve higher core and surface temperature, as it absorbs more current (the lower $I_{B_1}$ results in the higher $I_{cell_1}$).

In contrast, charging times are different for Cell 1 and Cell 2 when using the OCP-DCT scheme as demonstrated in \textcolor{black}{Fig.~\ref{fig.DCT25}}. As expected, Cell 2 (\textcolor{black}{at} higher initial SOC) is charged faster while both cells have the same rates of charging across the different  ambient temperatures. Once Cell 2 is fully charged at $t_{f_2}$, (i) $SOC_2$ is kept constant until Cell 1 reaches $SOC_{\textrm{target}}$ at $t_{f_1}$, (ii) the cell current is absorbed by the power units implementing the active balancing circuitry, leading to $I_0 = I_{B_2}$ over $t_{f_2}\leq t\leq t_{f_1}$, (iii) $V_{cell_2}$ drops at $t_{f_2}$ and remains constant over $t_{f_2}\leq t\leq t_{f_1}$, (iv) the core and surface temperatures of Cell 2 start decreasing at $t_{f_2}$, and (v) the rates of $L_{sei_2}$ and
$Q_2$ slow down after $t_{f_2}$. 
\textcolor{black}{The} figures also show that the OCP-DCT scheme reduces degradation gradient between cells (i.e., $L_{sei_2}(t_{f_2})-L_{sei_1}(t_{f_1})$) at all different ambient temperatures.

To compare the results of OCP-SCT and OCP-DCT, Table~\ref{table.Ex1} lists
quantitative comparisons between the two schemes at $T_{amb}=[15, 25, 35]^{\circ} \text{C}$. 
Referring to this table, under either OCP-SCT or OCP-DCT, when $T_{amb}$ increases, the following trends are inferred: (i) the SEI layer thickness variation of both cells increases, (ii) the capacity \textcolor{black}{loss} variation of 
\textcolor{black}{each} cell increases, and (iii) the charging time of 
\textcolor{black}{each} cell increases. 
According to this table, OCP-DCT decreases $\textrm{max}(\Delta L_{sei_1}, \Delta L_{sei_2})$ by 73\%, 40\%, and 40\%  
over OCP-SCT when $T_{amb}$ is set to $15^{\circ} \text{C}$, $25^{\circ} \text{C}$, and $35^{\circ} \text{C}$, respectively; this, in turn, leads to $\textrm{max}(\Delta Q_{1}, \Delta Q_{2})$ to be decreased by 72\%, 40\%, and 35\%, respectively when OCP-DCT is used. 
In terms of charging time, however, OCP-DCT increases $\textrm{max}(t_{f_1}, t_{f_2})$ by 40\%, 28\%, and 42\% over OCP-SCT when $T_{amb}$ is $15^{\circ} \text{C}$, $25^{\circ} \text{C}$, and $35^{\circ} \text{C}$, respectively.

\subsection{Robustness to initial SOC and SOH imbalances}
To further elaborate on the robustness of the proposed OCP-SCT and OCP-DCT schemes,
this section is devoted to perform multiple simulations at different ambient temperatures $[15, 25, 35] \ ^{\circ} C$ for initial values of $SOC$ and $SOH$ imbalance randomly taken from \textcolor{black}{uniform distributions}.

\subsubsection{Random initial SOC imbalance}
For each ambient temperature, $N_{sim}=20$ simulations are carried out where  initial SOCs are drawn from a uniform distribution over the interval $[0.2, \ 0.4]$. It can be seen from Fig.~\ref{fig.MC-SOC} that
under either scheme, when \textcolor{black}{the} ambient temperature increases, the maximum of SEI layer thickness variations of the
cells $\textrm{max}(\Delta L^*_{sei_1}, \Delta L^*_{sei_2})$ and the maximum
charging times of the cells $\textrm{max}(t^*_{f_1}, t^*_{f_2})$ increase as well.

Numerical results show that under \textcolor{black}{the} OCP-SCT scheme,  
\textcolor{black}{$\textrm{max} \ \underset{k_{N_{sim}}}{\textrm{max}}(\Delta L^*_{sei_1},\Delta L^*_{sei_2})=[0.08,9.62,57.15]\%$
and $\underset{k_{N_{sim}}}{\textrm{max}}(t^*_{f})=[510,596,717] \text{s}$} for $T_{amb}=[15, 25, 35]^{\circ} \text{C}$, where $k_{N_{sim}}=1,...,20$ is the $k_{N_{sim}}^{th}$ simulation. On the other hand, with OCP-DCT scheme and under different ambient temperatures,  
$\textrm{max} \ \underset{k_{N_{sim}}}{\textrm{max}}(\Delta L^*_{sei_1},\Delta L^*_{sei_2})=[0.04,6.22,38.64]\%$ and 
\textcolor{black}{$\textrm{max} \ \underset{k_{N_{sim}}}{\textrm{max}}(t^*_{f_1},t^*_{f_2})=[587,646,738]\text{s}$}. These findings are in agreement with our observations in Table~\ref{table.Ex1}, showing that the optimization under OCP-DCT scheme leads to the battery module with lower variation of SEI layer thickness and longer charging time regardless of the ambient temperature at which the simulation is performed.

\subsubsection{Random initial SOH imbalance}

In this experiment, $N_{sim}= 20$ simulations are run for  each ambient temperature for both control \textcolor{black}{schemes}, where in each simulation, initial $L_{sei}$ values are drawn from a uniform distribution over the interval $[4, 6]\times10^{-9} \text{m}$ to represent the SOH imbalance at the beginning of the battery life.
From Fig.~\ref{fig.MC-Lsei}, results reveal that with OCP-SCT scheme,  
$\textrm{max} \ \underset{k_{N_{sim}}}{\textrm{max}}(\Delta L^*_{sei_1},\Delta L^*_{sei_2})=[0.09,14.94,77.10]\%$
and $\underset{k_{MC}}{\textrm{max}}(t^*_{f})=[498,547,783] \ \text{s}$, and 
under OCP-DCT scheme,  
$\textrm{max} \ \underset{k_{N_{sim}}}{\textrm{max}}(\Delta L^*_{sei_1},\Delta L^*_{sei_2})=[0.06,9.98,57.87]\%$ and 
$\textrm{max} \ \underset{k_{N_{sim}}}{\textrm{max}}(t^*_{f_1},t^*_{f_2})=[546,620,815]\ \text{s}$ all for $T_{amb}=[15, 25, 35]^{\circ} \text{C}$. These results are in line with what we found from the case of initial SOC imbalance, showing that OCP-DCT scheme is able to mitigate the variation of SEI layer thickness at the cost of higher charging time 
\textcolor{black}{irrespective of the  ambient temperature}. In comparison with the case of initial SOC imbalance,
the simulations with initial SOH imbalance leads to the battery module with higher variation of SEI layer thickness at any ambient temperature used. 

\subsection{Pareto fronts: effect of trade-off coefficient $\alpha$} \label{subsec.pareto}

Recall that the optimization trade-off coefficient was picked to be $\alpha=0.5$ in the previous sections to study the \textit{balanced charging-degradation scenario}. However, this parameter could be varied to weigh more or less battery degradation over time of charge, given that the two costs have conflicting objectives. 
In this section, $\alpha$ is discretely sampled as $\alpha=\{0,\ 0.25,\ 0.5,\ 0.75, 1\}$ under which OCP-SCT and OCP-DCT schemes are run for different ambient temperatures when there is an initial SOC mismatch $SOC(0)=[0.2, 0.4]$. \textcolor{black}{Fig.~\ref{fig.Pareto}} shows that the maximum of SEI layer thickness variations of the
cells reduces as $\alpha$ decreases from 1 to 0 at any ambient temperature; the battery module ages less but takes more time for charging when we go from \textit{fast charging } to \textit{minimum degradation objective}. This is also supported by numerical results from which when $\alpha$ goes from 1 to 0, at $T_{amb}=[15, 25, 35]^{\circ} \text{C}$,
(i) under OCP-SCT, $\textrm{max} \ \underset{k_{N_{sim}}}{\textrm{max}}(\Delta L^*_{sei_1},\Delta L^*_{sei_2})$ decreases by $72\%$, $38\%$, and $23\%$, and $\underset{k_{N_{sim}}}{\textrm{max}}(t^*_{f})$ increases by $66\%$, $49\%$, and $75\%$, respectively; and  
(ii) under OCP-DCT, $\textrm{max} \ \underset{k_{N_{sim}}}{\textrm{max}}(\Delta L^*_{sei_1},\Delta L^*_{sei_2})$ decreases by $71\%$, $26\%$, and $27\%$, and $\textrm{max} \ \underset{k_{N_{sim}}}{\textrm{max}}(t^*_{f_1},t^*_{f_2})$ increases by $97\%$, $46\%$, and $53\%$, respectively. Once again, the Pareto fronts support our previously-claimed observations, showing that OCP-DCT scheme can reduce the battery degradation at any ambient temperature tested.  

\begin{figure*}[tp]
	\centering
	\subfigure[][]{\includegraphics[width=7.5cm]{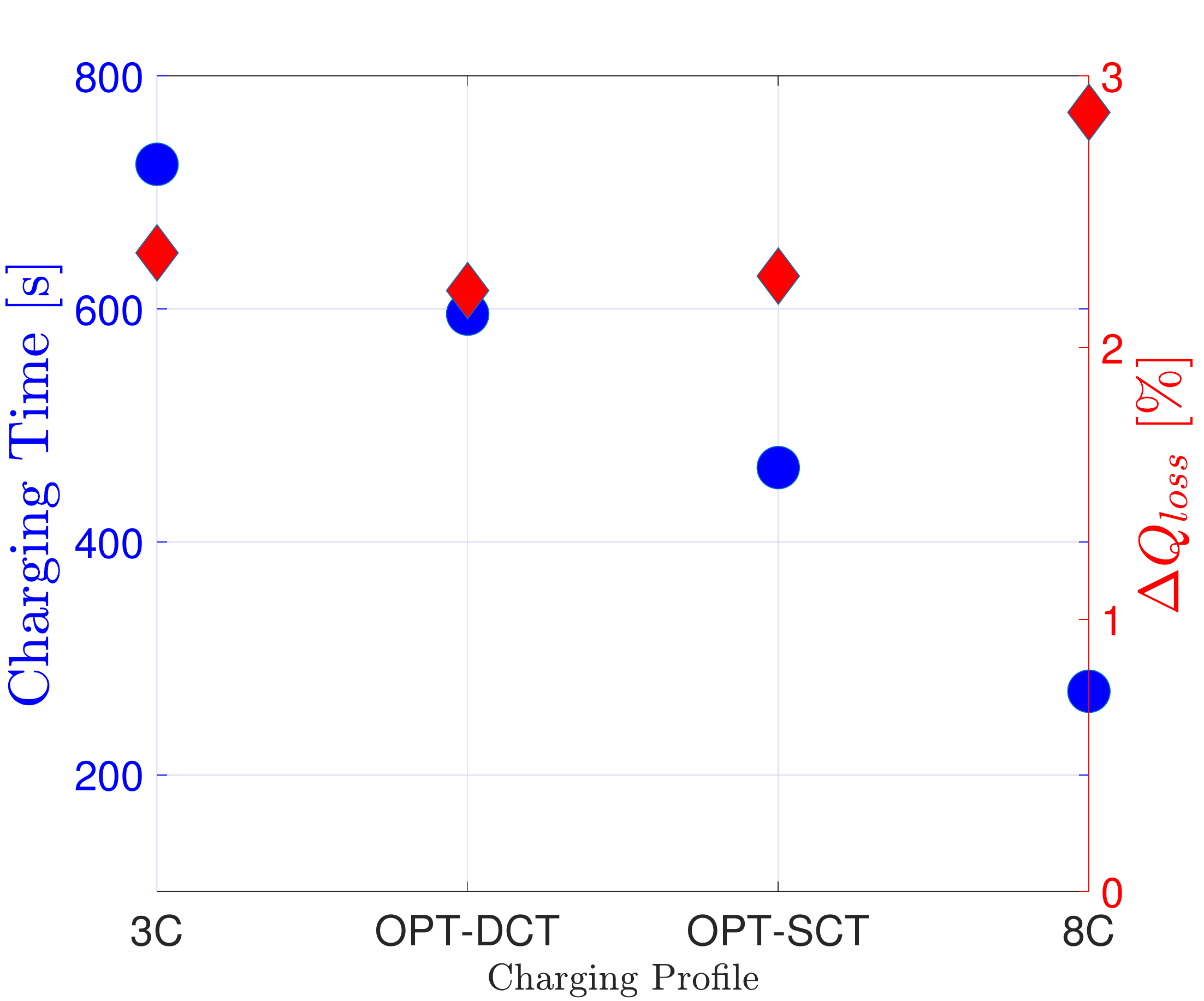}\label{fig.CC_Cell1}}	
	\subfigure[][]{\includegraphics[width=7.5cm]{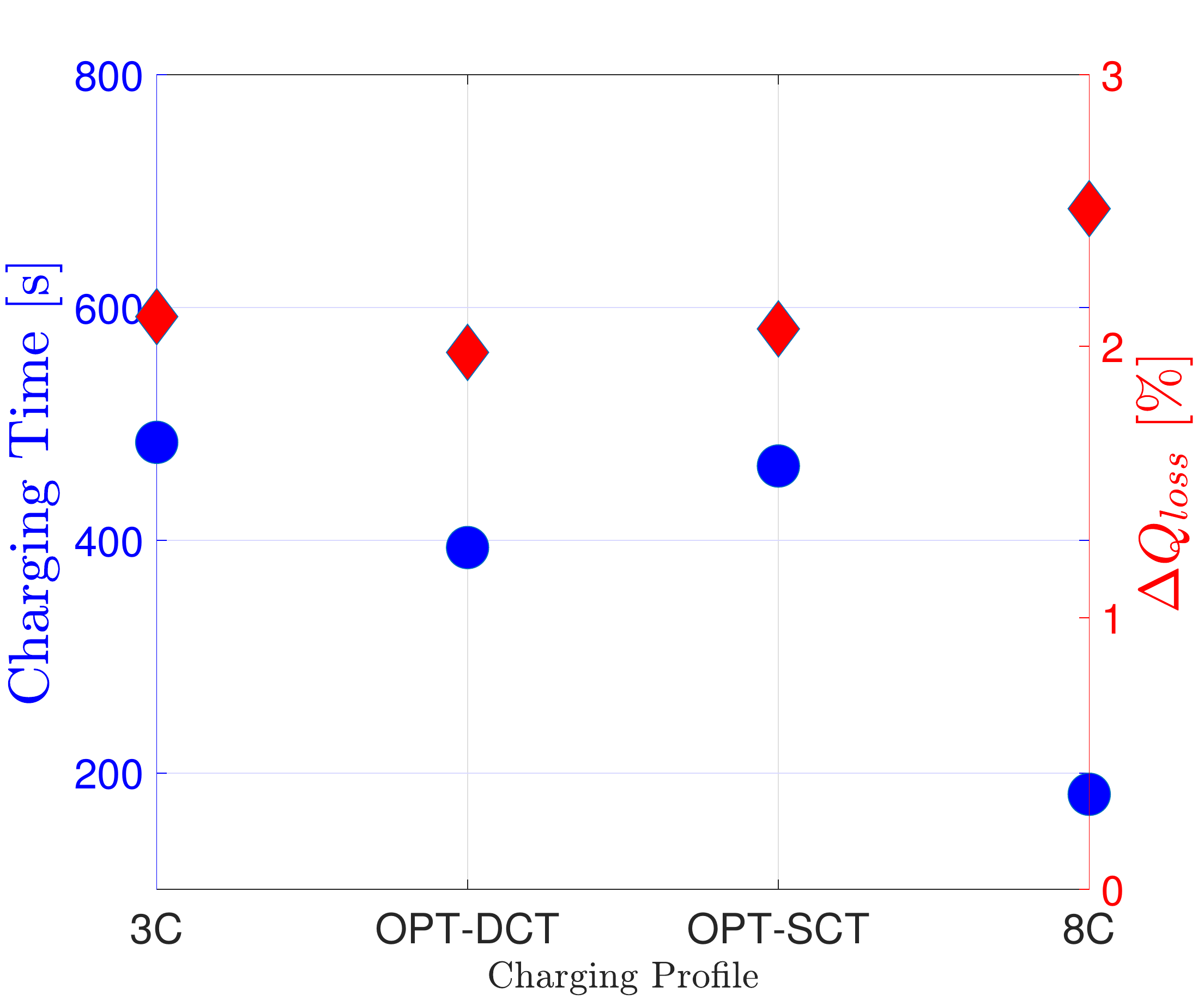}\label{fig.CC_Cell2}}
	\caption{\textcolor{black}{Charging time vs Capacity loss tradeoff for (a) Cell 1 and (b) Cell 2, with an initial mismatch of $SOC(0)=[0.2, 0.4]$, when subjected to 3C, OCP-SCT, and 8C profiles for 300 cycles, respectively.}}
\end{figure*}

\textcolor{black}{
\subsection{Comparison with conventional constant current profiles}
	To highlight the advantages and benefits of the proposed optimal controller, a comparison is made with the standard constant current (CC) charging profiles.\footnote{The CC-CV charging protocol is used in laboratory testing, whereas only CC - or its  variants- is used for in-vehicle charging.} Given that research efforts are underway to enable extreme fast charging, wherein the battery pack must be charged to $80\%$ of its capacity in 10-15 minutes \cite{tanim2019extreme}, it is reasonable to evaluate the performance of the proposed schemes against higher C-rates ($>$3C). The candidate CC charging profiles selected are 3C and 8C, which are the minimum and maximum permissible current magnitudes for the cell considered in this work. The two CC profiles along with the OCP-DCT and OCP-SCT profiles proposed in this work are applied to the battery module of two cells connected in series for 300 cycles each.\footnote{Note that one cycle is composed of the cells being charged from their initial SOC to the final SOC of 0.8.} An initial SOC imbalance of $SOC(0)=[0.2, 0.4]$ is assumed for the two cells in series. The performance of the series-connected cells under the four charging profiles [3C, OCP-DCT, OCP-SCT, 8C] are evaluated at an ambient temperature of $25^{o}C$ in terms of (a) charging time for the first cycle, and (b) capacity loss at the end of 300 cycles. \textcolor{black}{In this case, the capacity loss for a series-connected cell $k$ is defined as the percentage change in its capacity at the end of 300 cycles, with respect to the nominal capacity, given by $\Delta Q_{loss}^{k}=\frac{Q_{nom}^{k} - Q_{300 cycles}^{k}}{Q_{nom}^{k}}\times 100\%$. This study intends to demonstrate the health savings each charging strategy offers, in terms of retained capacity over multiple charging cycles}. 
%
\\ \\
	In Fig.~\ref{fig.CC_Cell1}, we plot the charging time (blue circle) and capacity loss at the end of 300 cycles (red triangle) of Cell 1 for the charging profiles 3C, OPT-DCT, OPT-SCT and 8C, respectively. It is noticed that the charging time reduces as the C-rate increases from 3C to 8C, and as expected, the amount of degradation has the opposite trend wherein as the C-rate increases, the observed capacity loss is higher. However, interestingly, the capacity loss observed for the OCP-SCT and OCP-DCT profile is lower than 8C and slightly lower than 3C, thereby providing a balanced charging-degradation solution. This indicates that the proposed optimal control profile results in not only minimum degradation compared to both 3C and 8C profiles, but also provides a good trade-off in charging time between the two extremes of 3C and 8C. Similar trends are also observed in the capacity loss of Cell 2 in Fig.~\ref{fig.CC_Cell2} for all charging profiles. Cell 2 has a higher initial SOC, and hence its charging time for the OCP-DCT profile is 
	\textcolor{black}{shorter} because the scheme allows for different charging times of the cells to account for heterogeneous initial conditions, whereas the charging time of Cell 2 is same as that of Cell 1 for the OPT-SCT profile. The results validate that the OCP-DCT and OCP-SCT profile outperform the standard CC profiles by providing a balanced trade-off between fast charging and minimum degradation.  
	Note that these results are simulated for 300 charging cycles, however, each cycle only consists of a SOC window from 0.2 or 0.4 to 0.8 (depending on initial SOC of cells in the module). It follows that as the battery ages and undergoes long-term cycling, the trends and savings, in terms of capacity, will be more pronounced, thereby highlighting the advantages of the proposed optimal controller.} 


\section{Conclusion and Discussion}\label{section.Conc}

\subsection{Conclusion}

This paper formulated a multi-objective fast charging-minimum degradation OCP for battery modules with $N_{cell}$ series-connected cells with an active balancing circuitry.  A surrogate model was proposed to mitigate computational burden associated with the multi-time scale nature of the cell dynamics as well as the large scale nature of LIB modules. Two different OCPs were suggested: OCP-SCT and OCP-DCT. Simulation studies were carried out on a battery module with two series-connected cells in the presence of initial SOC and SOH imbalances under different ambient temperatures. Results demonstrated that under both schemes outperform standard CC charging profiles, and degradation and charging time increase as ambient temperature increases. Our findings showed that OCP-DCT provides more flexibility \textcolor{black}{to handle heterogeneities}
among the cells in terms of obtaining a more uniform degradation among the cells, hence  leading to a longer utilization 
of the module.

\textcolor{black}{In the future, the optimal control of series-connected modules during discharging will be investigated. In the discharging case, the module current $I_0$ is fixed as per the current/power demand requested by the user or the application, resulting in one less degree of freedom and optimization variable. However, the objective functions will need to be modified according to the discharging scenarios (for instance: charging time objective function is not valid). Having said that, the framework proposed in this paper, which consists of using the direct collocation approach to transcribe the OCP into a NLP problem by parameterization of the system states and input, will remain the same.} \textcolor{black}{It is worth mentioning that as the number of cells in series increase, the computational burden of solving the optimal control problem will be higher. To that end, the proposed optimal controller is more suited for offline simulations of series-connected cells to generate solutions, trajectories, or reference surface maps, to aid our understanding of the optimal split under different conditions, and identify critical conditions or faults. The results from the offline simulations can be used in the form of look-up tables or maps for reference tracking during real-time applications (with reduced-order models) in a resource-constrained on-board hardware.}

\subsection{Discussion: impact of our work}
The adoption of an effective active balancing hardware in a battery pack holds the potential to address the issue of  guaranteeing longer ($>$8 years) life when  used in EV applications.
\textcolor{black}{Currently, cell balancing via shunt resistors is widely used in the industry. Since the proposed optimal controller is applicable to any general active balancing hardware (either shunt resistors, transistors, DC/DC converters), it is easier to adopt and it can be immediately deployed without adding additional hardware costs.}\\
In a series-connected module, the capacity of the module is defined by the weakest (most aged) cell. Heterogeneity among cells, if not embraced, will result in some cells to be overly used over time thus creating a fragile (age-wise) link in the module. The ability to control each single cell while acknowledging their initial states, health and manufacturing characteristics will result in a module/pack with uniform characteristics and performance. 
In the quest for solutions that provide longer battery life capability, among discovering new materials  and proposing novel manufacturing processes, the system level solution explored in this paper positions itself as an easily deployable method for targeted applications.

\begin{table}[t]
	\centering
	\begin{tabular}{ll}
		\hline
		\textbf{Acronyms} & \\
		BP  & Break point\\
		CCCV & Constant-current constant-voltage \\
		CP  & Collocation point \\
		DAE &Differential algebraic equation \\
		DOD & Depth of discharge \\
		DCT& Different charging time  \\
		FDM&  Finite difference method \\
		GQF  & Gaussian quadrature formula \\
		IPOPT  & Interior point optimizer \\
		KKT  & Karush-Kuhn-Tucker \\
		LIB  & Lithium-ion battery\\
		NLP  & Nonlinear programming\\
		NMC  & Nickel–manganese–cobalt \\
		OCP  & Optimal control problem \\
		ODE & Ordinary differential equation\\
		PDE  & Partial differential equation\\
		SCT& Same charging time  \\
		SEI& Solid electrolyte interphase \\
		SOC& State of charge\\
		SOH& State of health\\
		SPM & Single particle model  \\
		\hline
	\end{tabular}
\end{table}
\section{Acknowledgements}\label{sec:Ack}
The authors are grateful to LG Chem (now LG ES) for the financial support, and they would like to thank Dr. Won Tae Joe and Dr. Yohwan Choi for providing invaluable guidance for this work.

\begin{table*}[!]
	\renewcommand{\arraystretch}{2}
	\caption{Nomenclature.}
	\centering
	\label{table.ESPM_nomenclature}
	\centering
	\resizebox{18cm}{!}{	
		\begin{tabular}{l l l l l l}
			\hline\hline \\[-6mm]
			$c_{s,j}$ & Concentration in solid phase [\text{mol/m$^3$}]
			&$c_{e}$ & Concentration in electrolyte phase [\text{mol/m$^3$}]
			&$c_{solv}$ & Solvent concentration [\text{mol/m$^3$}] \\
			$T_c$ & Cell core temperature [\text{K}]
			&$T_s$ & Cell surface temperature [\text{K}]
			& $L_{sei}$ & SEI layer thickness [\text{m}] \\
			$Q$ & Cell capacity [\text{Ah}]
			&$I_{cell}$ & Cell current [\text{A}]
			&$\eta_j$ & Overpotential [\text{V}] \\
			$i_{0,j}$ & Exchange \textcolor{black}{current density} [\text{A/m$^2$}]	
			&$U_{j}$ & Open circuit potential (electrode) [\text{V}]
			&$V_{oc}$ & Open circuit \textcolor{black}{voltage} (cell) [\text{V}] \\
			$i_s$ & Side reaction current density [\text{A/m$^3$}]
			&$D_{s,j}$ & Solid phase diffusion [\text{m$^{2}$/s}]
			& $R_{s,j}$ & Particle radius [\text{m}] \\
			$a_{s,j}$ & Specific interfacial surface area [\text{m$^{-1}$}] 
			&$A$ & Cell cross sectional area [\text{m$^{2}$}]
			& $L_{j}$ & Domain thickness [\text{m}] \\
			$F$ & Faraday's constant [\text{C/mol}] 
			&$c_{s,j}^{max}$ & Maximum electrode concentration [\text{mol/m$^3$}]
			& $\kappa^{eff}_{j}$ & Effective electrolyte conductivity [\text{S/m}] \\
			$k_{j}$ & Reaction rate constant [\text{m$^{2.5}$/s-mol$^{0.5}$}]
			& $R_{l}$ & Lumped contact resistance [\text{$\Omega$}] 
			&$R_{el}$ & Electrolyte resistance [$\Omega$] \\
			$R_{sei}$ & SEI layer resistance [\text{$\Omega$}]
			& $R_{g}$ & Universal gas constant [\text{J/mol-K}]	
			& $D_{solv}$ & Solvent diffusion in SEI layer [\text{m$^{2}$/s}] \\
			$\epsilon_{sei}$ & SEI \textcolor{black}{layer} porosity
			& $\rho_{sei}$ & SEI layer density [\text{kg/m$^3$}]
			&$\kappa_{sei}$ & SEI layer ionic conductivity [\text{S/m}] \\
			$c_{solv}$ & Solvent concentration [\text{mol/m$^3$}]
			&$M_{sei}$ & Molar mass of SEI layer [\text{kg/mol}]	
			&$\beta$ & Side \textcolor{black}{reaction} charge transfer coefficient \\
			$C_s$ & Heat \textcolor{black}{capacity} of cell surface [\text{J/K}] 
			&$C_c$ & Heat \textcolor{black}{capacity} of cell core [\text{J/K}] 
			&$R_c$ & Conductive resistance - core/surface [\text{K/W}] \\	
			$R_u$ & Convective resistance - surface/surroundings [\text{K/W}] 
			& $T_{amb}$ & Ambient temperature [\text{K}]
			&$N_{r,j}$ & Number of radial discretization points \\
			$N_{sei}$ & Number of SEI layer discretization points
			&$c^{surf}_{s,j}$ & Surface concentration in solid phase [\text{mol/m$^3$}]
			&$c^{surf}_{solv}$ & Surface solvent concentration [\text{mol/m$^3$}] \\
			$c^{avg}_{e}$ & Average electrolyte concentration [\text{mol/m$^3$}]
			&$\theta^{surf}_j$ & surface stoichiometry in solid phase 
			&$c^{bulk}_{s,j}$ & Bulk concentration [\text{mol/m$^3$}] \\
			$E_{a,\varphi}$ & Activation energy [J/mol]
			&$\theta^j_{0\%}$ & Reference stoichiometry ratio at 0\% SOC 
			&$\theta^j_{100\%}$ & Reference stoichiometry ratio at 100\% SOC  \\
			$k_f$ & Solvent \textcolor{black}{reduction} rate constant [\text{mol$^{-2}$s$^{-1}$}]
			&$c^*_{solv}$ & Optimal solvent concentration [\text{mol/m$^3$}]
			&$\kappa_{sei}$ & SEI layer ionic conductivity [S/m] \\ 
			$I_{cell}$ & Cell current [A] 
			& $V_{cell}$ & Cell voltage [V]
			& $T_{c,ref}$ & Reference core temperature [K] \\
			$\epsilon_{e,j}$ & Electrolyte porosity  
			&$\Phi_{s,n}$ & Anode surface potential [V]
			& $U_{s}$ & Solvent reduction potential [V] \\
			$r$ & Radial coordinate
			& $R_m$ & Cell-to-cell heat transfer resistance [K/W]
			& $N_{cell}$ & Number of cells  \\
			$x$ & State vector
			& $I_0$ & Module current [A]
			& $I_B$ & Balancing current [A] \\
			$X$ & Optimization variable vector 
			& $t_f$ & Charging time [s]
			& $N_s$ & Number of states  \\
			$N_{BP}$ & Number of break points 
			& $Q_{nom}$ & Nominal capacity [Ah]
			& $\alpha, \beta_1, \beta_2, \beta_3$ & Optimization parameters  \\
			$s_p$ & Smoothness degree 
			& $d_p$ & Polynomial order 
			& $P$ & Free parameter set  \\
			$\mu_1,\mu_2$ & KKT multipliers 
			& $\mathcal{B}_{p,q}$ & B-Spline 
			& $\omega_{p,q}$ & Free parameters of optimization variables \\
			$\phi$ & Thermal diffusivity
			& $R_{cell}$ & Radius of a cylindrical 18650 cell [m]
			\\[0.5mm]
			\hline
		\end{tabular}
	}
\end{table*}

\bibliographystyle{ieeetr}
\bibliography{myrefs2}
\end{document}